\documentclass[aps,prl,preprint,groupedaddress,showpacs]{revtex4-2}

\usepackage{graphicx}
\usepackage{color} 
\usepackage{subcaption} 

\begin{document}


\title{Strain hardening by sediment transport}



\author{Fernando D. C\'u\~nez}
 \email{david.cunez@hotmail.com}
\author{Erick M. Franklin}
 \email{erick.franklin@unicamp.br}
\affiliation{%
School of Mechanical Engineering, UNICAMP - University of Campinas,\\
Rua Mendeleyev, 200, Campinas, SP, Brazil\\
}%

\author{Morgane Houssais}
 \email{housais.morgane@gmail.com}
\affiliation{Levich Institute, City College of New York - CUNY\\
140th Street and Convent Avenue, NewYork, NY 10031, USA\\}

\author{Paulo Arratia}
 \email{parratia@seas.upenn.edu}
\affiliation{Mechanical Engineering and Applied Mechanics, University of Pennsylvania,\\
Philadelphia, 19104, Pennsylvania, USA\\}

\author{Douglas J. Jerolmack}
 \email{sediment@sas.upenn.edu}
 \thanks{Corresponding author}
\affiliation{Department of Earth and Environmental Science, and Mechanical Engineering and Applied Mechanics, University of Pennsylvania,\\
Philadelphia, 19104, Pennsylvania, USA\\}



\date{\today}

\begin{abstract}
The critical fluid-shear stress for the onset of sediment transport, $\theta_c$, varies with the history of applied shear. This effect has been attributed to compaction, but the role of shear jamming is unexplored. We examine the response of a granular bed to fluid-shear stress cycles of varying magnitude and direction, and determine isotropic and anisotropic contributions. Creep and bed-load transport result in direction-dependent strain hardening for $\theta/\theta_c < 4$. Dilation-induced weakening, and memory loss, occurs for larger stresses that fluidize the bed. Our findings provide a granular explanation for the formation and breakup of hard packed river-bed 'armor'.
\end{abstract}


\maketitle


The entrainment of granular materials by fluid shear acts to shape diverse landscapes on Earth and other planets \cite{jerolmack2019viewing}. While decades of research mostly examined the influence of fluid-flow structures and turbulence, a recent focus on granular rheology and structure has spurred progress in determining the roles of collisions and viscous dissipation in sediment transport dynamics and rates \cite{houssais2016rheology, clark2015onset, Houssais_1, maurin2016dense, pahtz2018universal, allen2017depth, Pahtz_4}. A central challenge that emerges is understanding the nature of the threshold of motion itself. In liquid-driven flows such as rivers, `bed-load' transport is often envisioned as a thin layer of surface grains moving over a static or `jammed' sediment bed \cite{Raudkivi_1, Charru_1}. In this context, the onset of bed-load transport is typically defined by the dimensionless critical Shields stress, $\theta_c$, which is a threshold value of the Shields number, $\theta \equiv \tau / ((\rho_s - \rho)gD)$, where $\tau$, $\rho_s$, $\rho$, $g$ and $D$ are the fluid-shear stress, particle density, fluid density, gravity and grain size. Clark et al. \cite{clark2017} offered a new interpretation of $\theta_c$ as the stress at which moving grains can no longer find a stable (static) configuration; this opens the possibility of relating the susceptibility of fluid entrainment to granular confinement. On the other hand, Houssais et al. \cite{Houssais_1} observed that grains beneath the threshold for bed-load motion were not static, but rather exhibited a slow and erratic creeping motion with caged dynamics. In their laminar flow experiments, $\theta_c$ coincided with a transition from creep to a dense-granular flow on the surface.

Temporal variation in the entrainment threshold is commonly observed in natural rivers; $\theta_c$ changes as a function of the history of fluid stress \cite{turowski2011start, Masteller_2, pretzlav2020smartrock}, causing hysteresis in observed sediment transport rates through a flood \cite{mao2014bedload, roth2017bed, pretzlav2020smartrock}. Experiments indicate that these transient dynamics arise due to the formation and breakup of river-bed `armor' \cite{Masteller_1, hassanexperimental}. Armoring has been proposed to arise from vertical grain-size segregation in polydisperse river beds \cite{dietrich1989sediment, blom2006vertical, ferdowsi2017river}, and by formation of a hard-packed `pavement' surface \cite{parker1982model, Charru_1, yager2018resistance, prancevic2015particle, Masteller_1}. In this study we consider the latter, in which the increased resistance of river-bed sediments to fluid entrainment may be considered a form of strain hardening. Charru et al. \cite{Charru_1} observed a steady decline in bed-load transport rate --- over several days --- under constant $\theta$, which they attributed to an increase in effective $\theta_c$ by compaction of the near-bed region. Even subcritical ($\theta < \theta_c$) flows cause armoring due to granular creep \cite{Masteller_1, Allen_2}, which can only be broken up by flows far exceeding critical ($\theta \gg \theta_c $ ). Using an experimental setup similar to ours, Allen and Kudrolli \citep{Allen_2} focused on the transient response of a sediment bed to sub-critical fluid shear. They observed that that creep rates diminished over time, while the granular packing fraction $\phi$ of the sediment bed increased toward the volume fraction associated with the glass transition, $\phi_{RCP} \simeq 0.64$. Moreover, they showed that the value $\theta_c$ needed to initiate bed load systematically increased with $\phi$. 

The findings above are consistent with the general picture that the rigidity of granular materials is primarily controlled by packing fraction or, more specifically, the distance from the jamming/glass transition characterized by $\phi_{RCP} - \phi$ \cite{Liu, keys2007measurement, boyer2011unifying, jerolmack2019viewing}. It is now known, however, that strain hardening may also occur without any change in volume. This ``shear jamming'' arises from the development of a granular fabric that is oriented to resist an applied, directional boundary shear  \cite{Cates, Majmudar, bi2011jamming, behringer2018physics}. Like compaction, the resulting strain hardening produces a history-dependent memory in the mechanical stability of the granular pack \cite{Keim}. But unlike compaction, this memory is stored in anisotropic structures that are fragile when subject to changes in shear direction \cite{bi2011jamming}. Shear jamming and compaction may operate in tandem; it is well known that shearing or shaking in multiple directions allows more compaction than one direction alone (see, e.g., \cite{yang2021evolution}) --- presumably because the suppression/destruction of anisotropic structures allows the bed to access a higher $\phi$ configuration 

We expect that fluid-sheared granular beds experience both isotropic (compaction) and anisotropic (shear jamming) strain hardening, due to the free-surface condition and directionality of applied shear. 
Sediment transport studies have not, however, considered anisotropic contributions. Moreover, the contributions of creep (sub-critical) vs. bed-load (super-critical) transport to strain hardening have not been isolated, and the conditions that lead to armor breakup and erasure of memory have not been characterized. In this Letter we address these issues with laminar sediment transport experiments, in which a sedimented bed of grains was subject to fluid shear-stress cycles of 10-min duration, with varying magnitude and direction. The device and materials are completely described in previous work \cite{Houssais_1}. Briefly: A bed (20-22 mm thick) of monodisperse acrylic beads (diameter, $d = 1.5$ mm; density, $\rho_s = 1190$ kg/m$^3$) was submerged in a refractive-index matched oil (kinematic viscosity, $\mu = 7.2 \times 10^{-2}$ Pa s; density, $\rho = 1050$ kg/m$^3$) inside an annular flume (Fig. \ref{fig_setup}(a)). A rotating lid was used to achieve fluid shear rates of 2.9 s$^{-1}$ $\leq$ $\dot{\gamma}$ $\leq$ 14.5 s$^{-1}$, which corresponds to Shields numbers of 0.1 $\leq$ $\theta$ $\leq$ 0.5 where $\tau$ = $\mu \dot{\gamma}$. Channel and grain scale Reynolds numbers, $Re$ = $\rho U_f h_f/\mu$ and $Re_s$ = $\rho \dot{\gamma} d^2/\mu$ respectively, were within 0.7 $\leq$ $Re$ $\leq$ 3.5 and 0.077 $\leq$ $Re_s$ $\leq$ 0.384, where $U_f$ is the lid velocity at the channel centerline and $h_f$ is the gap distance between the lid and bed surface. The preparation protocol involved almost complete suspension of the granular bed by pre-shearing for 5 min at $\theta$ = 6, followed by 5 minutes of quiescence to allow settling. From our previous work we choose a reference critical Shields stress of $\theta_{c} = 0.1$ \cite{houssais2016rheology}. Images captured at 60 Hz in the illuminated mid-channel cross section (Fig. \ref{fig_setup}(b)) allowed determination of vertical profiles of granular velocity, strain rate and concentration, averaged over each 600-s stress cycle and horizontally in the $x$ direction. Average velocity ($\langle V \rangle$) was determined from standard Particle Image Velocimetry (PIV) techniques, and average concentration ($\langle C \rangle$) was estimated from image filtering and thresholding \cite{Supplemental}. Strain was determined in the following manner: from one image to the next, a mobility matrix $M(x,z)$ was measured as the intensity difference at each pixel and its dimensionless equivalent, averaged in time and the $x$ direction, $m^*(z)$ (see Fig. \ref{fig_setup}, and the Supplemental Material \cite{Supplemental}); an effective strain was estimated at each elevation as $\epsilon(z) = \langle C_{\rm sat}\rangle /\langle C \rangle (z) \int^{z+\Delta z}_{z} m^* dy$.

\begin{figure}
\centering
\includegraphics[width=0.99\linewidth]{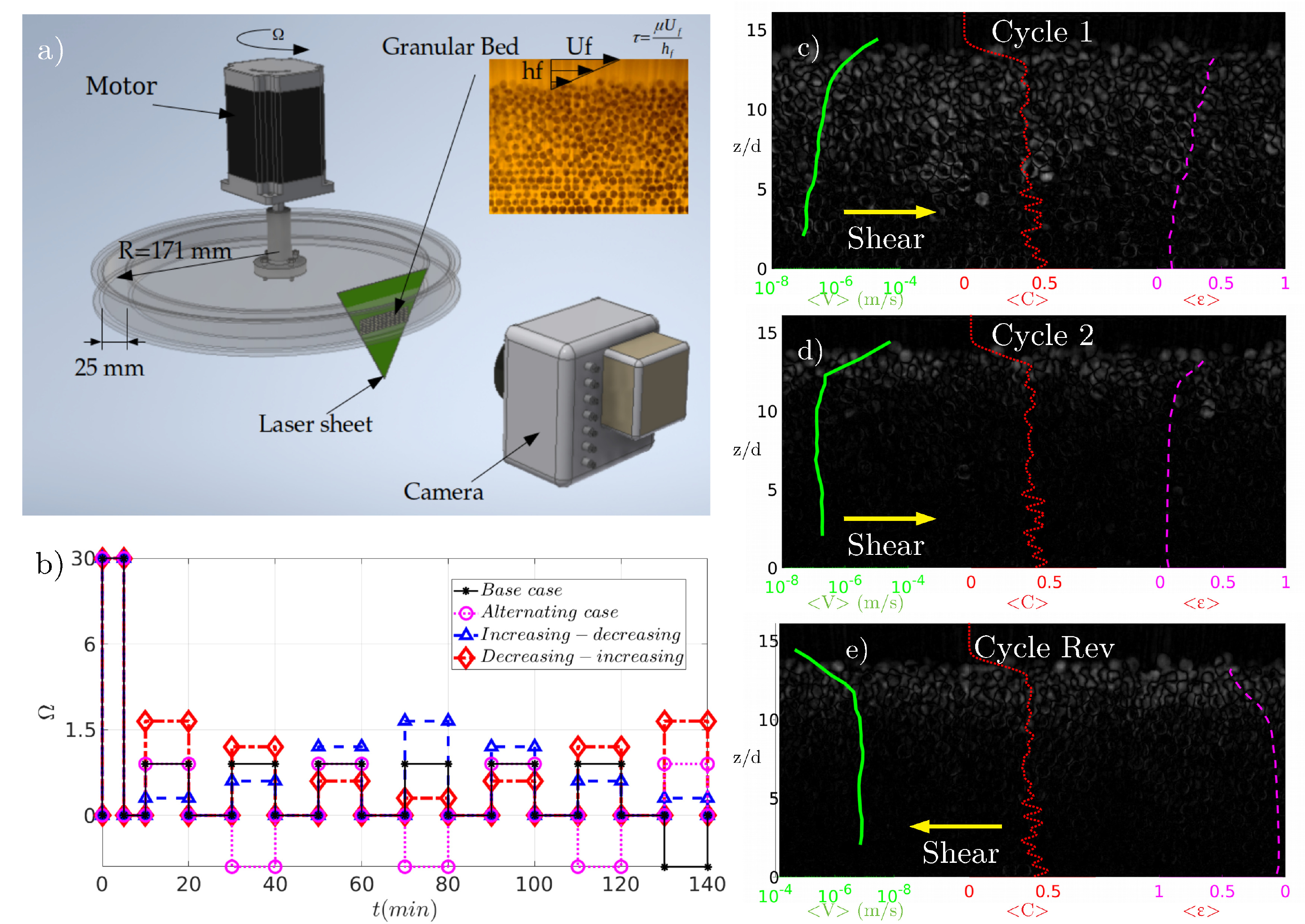}
\caption{(a) Schematic of the experimental setup. (b) Image of the laser-illuminated plane. (c-e) Strain (background in gray scale, where the light spots correspond to higher strain values), longitudinal and time-averaged particle velocities ($\langle V \rangle$, green line) and concentrations ($\langle C \rangle$, red-dotted lines), and longitudinal-averaged strains (magenta-dashed lines) for indicated cycles in the ``base case'' experiment.}
	\label{fig_setup}
\end{figure}

\begin{figure}
	\begin{minipage}[c]{0.49\linewidth}
		\begin{center}
			\includegraphics[width=.99\linewidth]{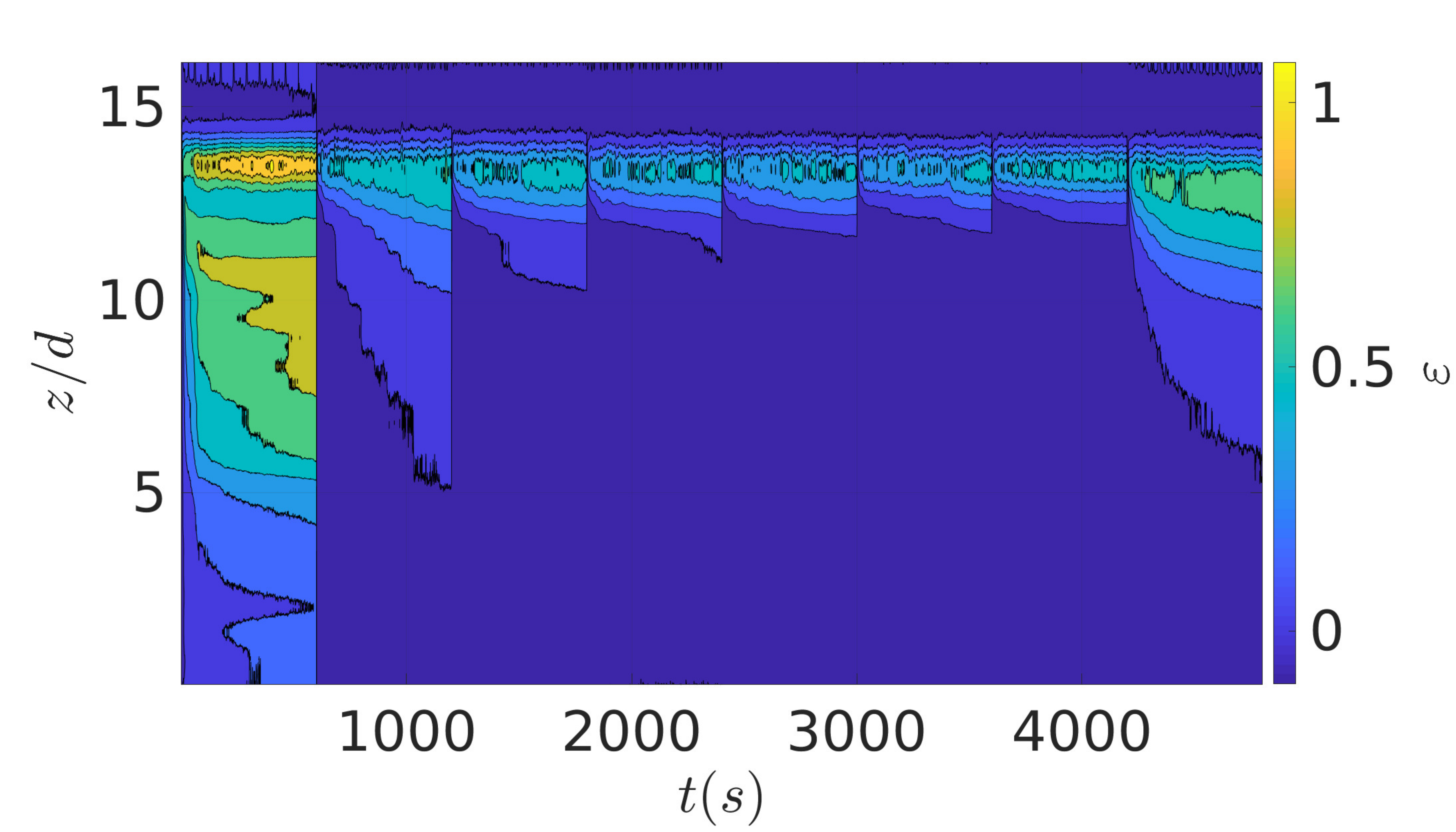}\\
			(a)
		\end{center}
	\end{minipage}
	\begin{minipage}[c]{0.49\linewidth}
		\begin{center}
			\includegraphics[width=.99\linewidth]{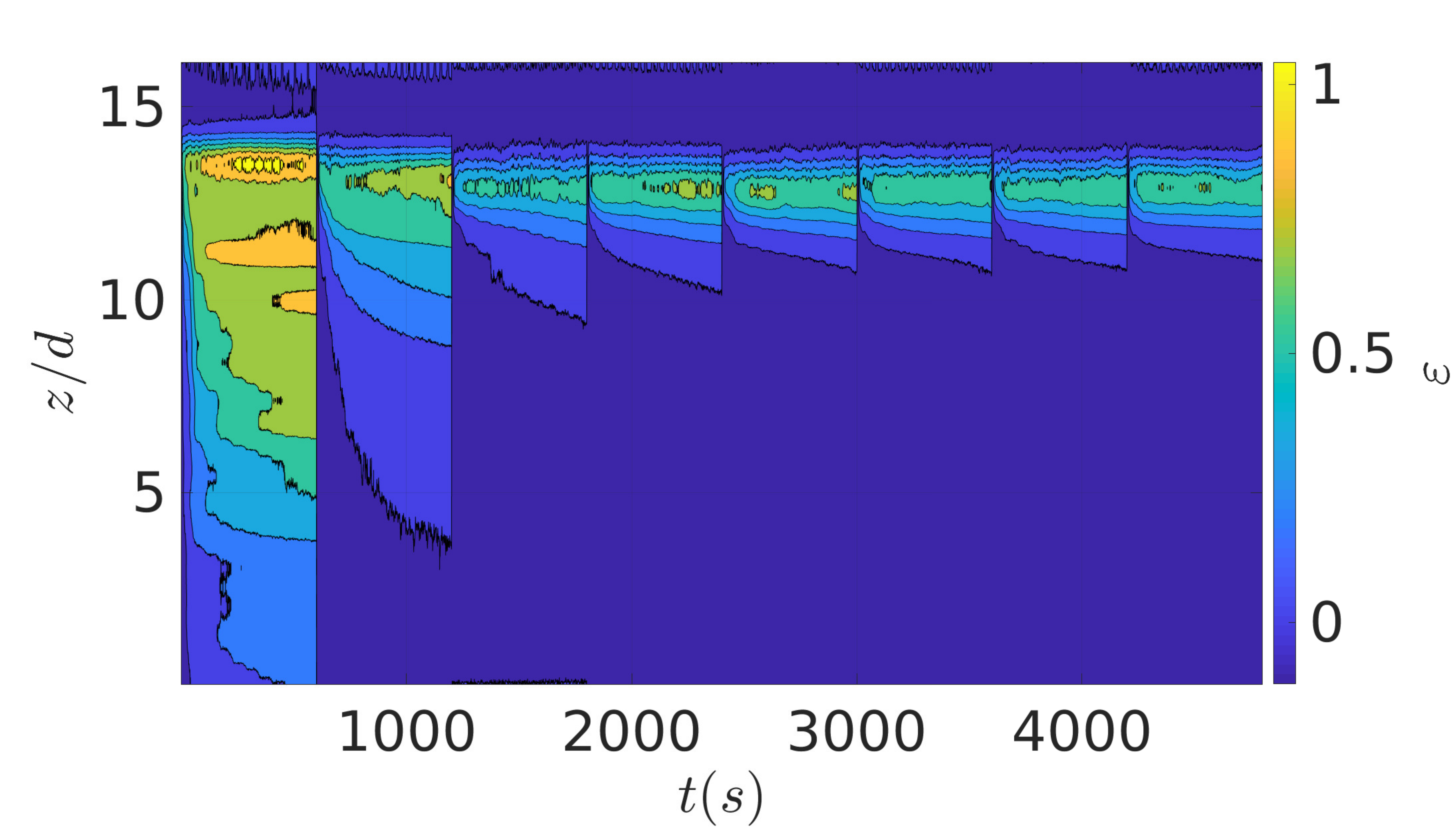}\\
			(b)
		\end{center}
	\end{minipage}
	\\
	\begin{minipage}[c]{0.49\linewidth}
		\begin{center}
			\includegraphics[width=.9\linewidth]{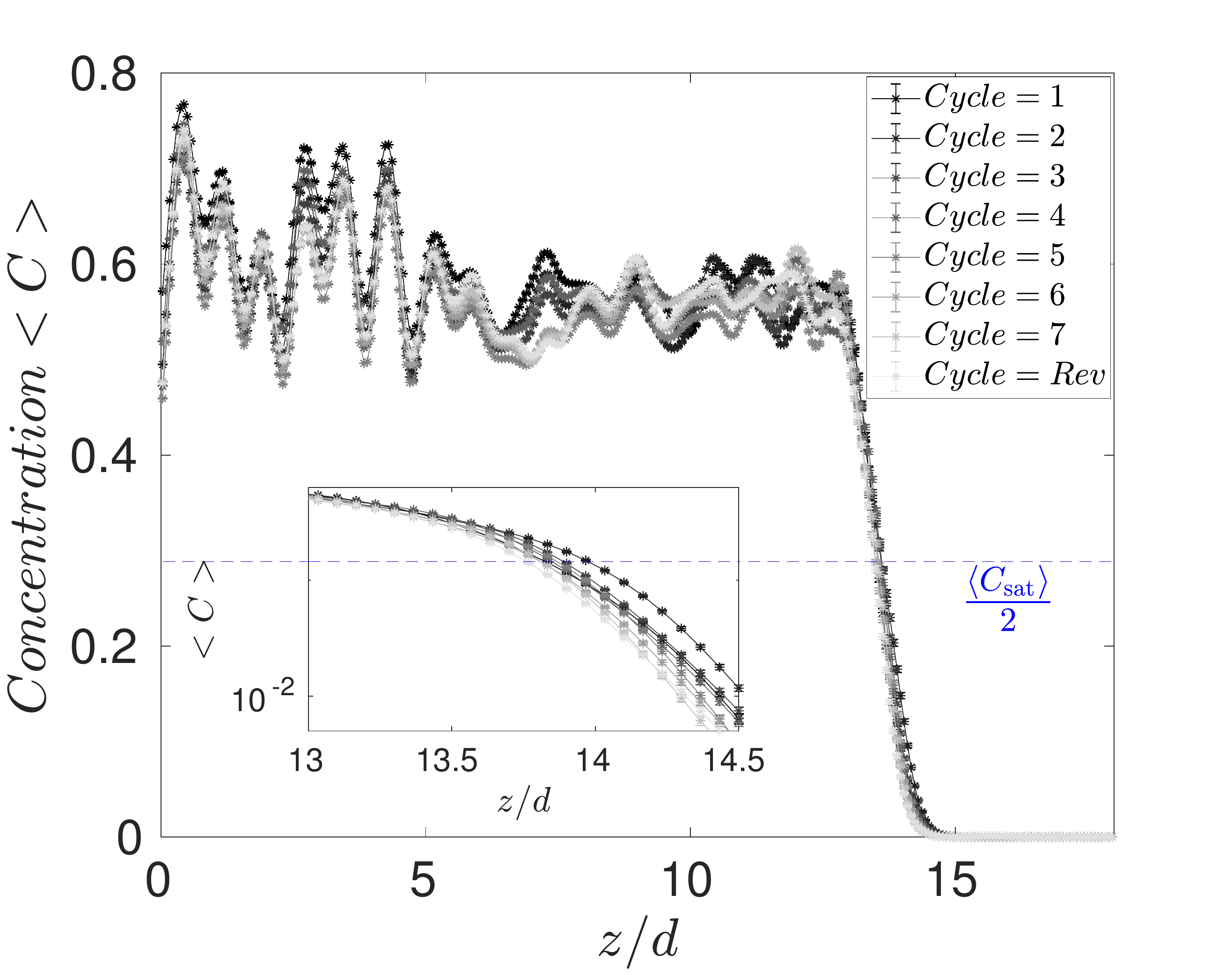}\\
			(c)
		\end{center}
	\end{minipage}
	\begin{minipage}[c]{0.49\linewidth}
		\begin{center}
			\includegraphics[width=.9\linewidth]{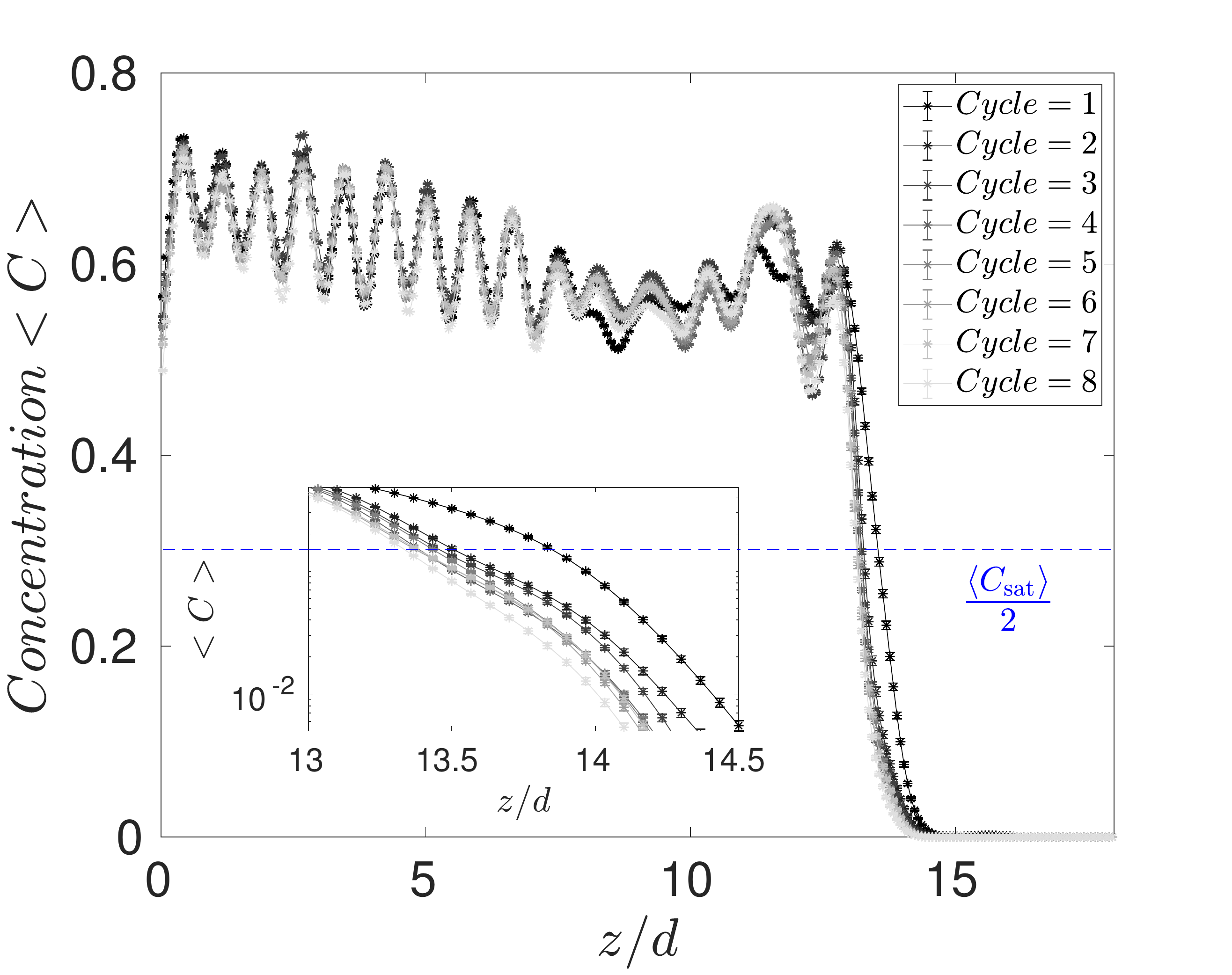}\\
			(d)
		\end{center}
	\end{minipage}
	\hfill
	\caption{$M$ matrix representing local particles' mobility intensity for (a) the ``base case'' with reversal at the last cycle, and (b) ``alternating case'' where stress is reversed for each cycle, and variation from blue (darker shades in grayscale) to yellow (brighter shades in grayscale) corresponds to increasing in mobility. Time averaged vertical concentration $\left<C(z)\right>$ at each cycle for (c) the ``base case'' with reversal at the last cycle, and (d) ``alternating case'' where stress is reversed for each cycle. $\theta$ = 2$\theta_{c}$ for all cycles, with gray levels scaling with time.}
	\label{displacements1}
\end{figure}

\begin{figure}
\begin{minipage}[c]{0.32\linewidth}
	\begin{center}
		\includegraphics[width=\linewidth]{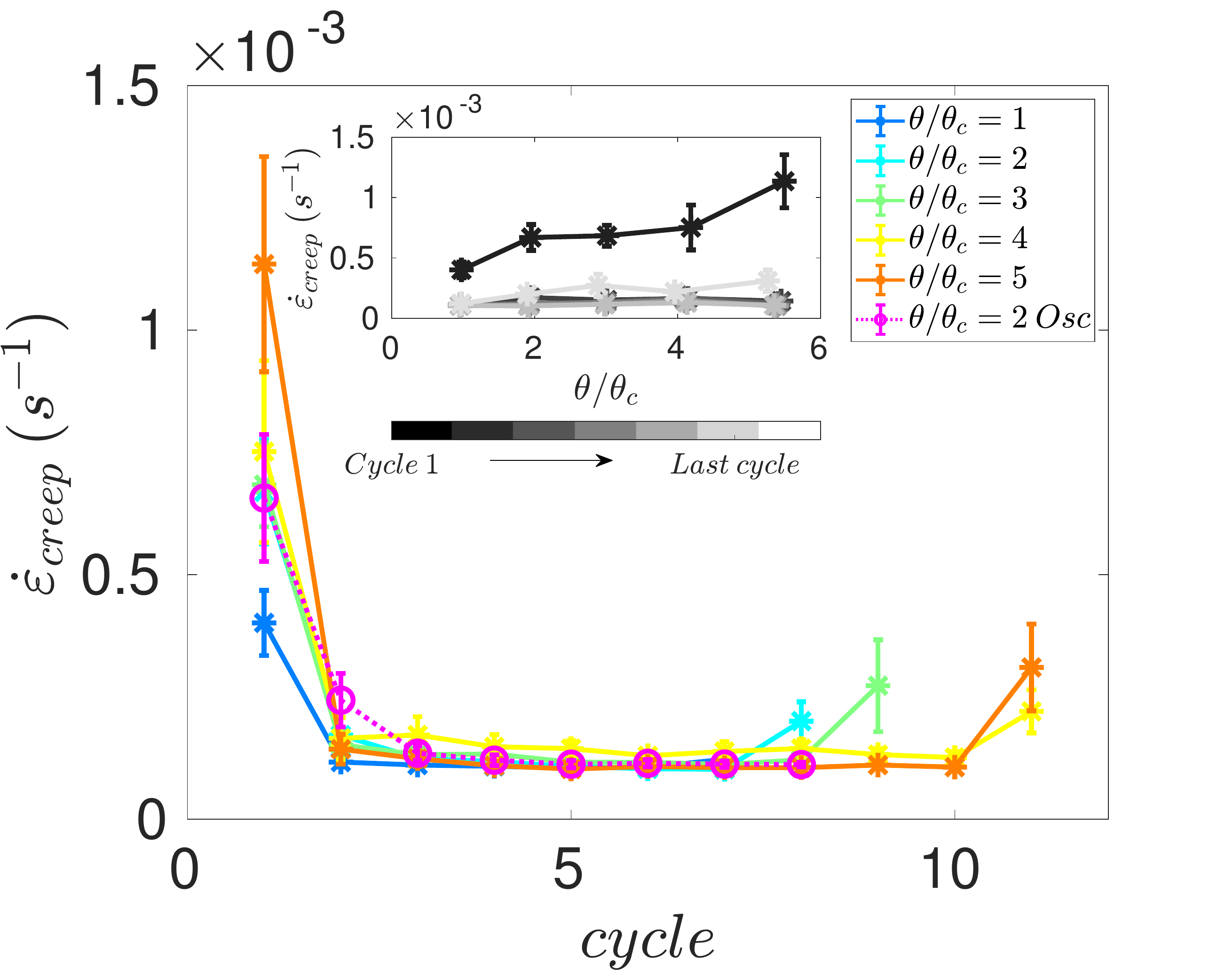}\\
		(a)
	\end{center}
\end{minipage}
\begin{minipage}[c]{0.32\linewidth}
	\begin{center}
		\includegraphics[width=\linewidth]{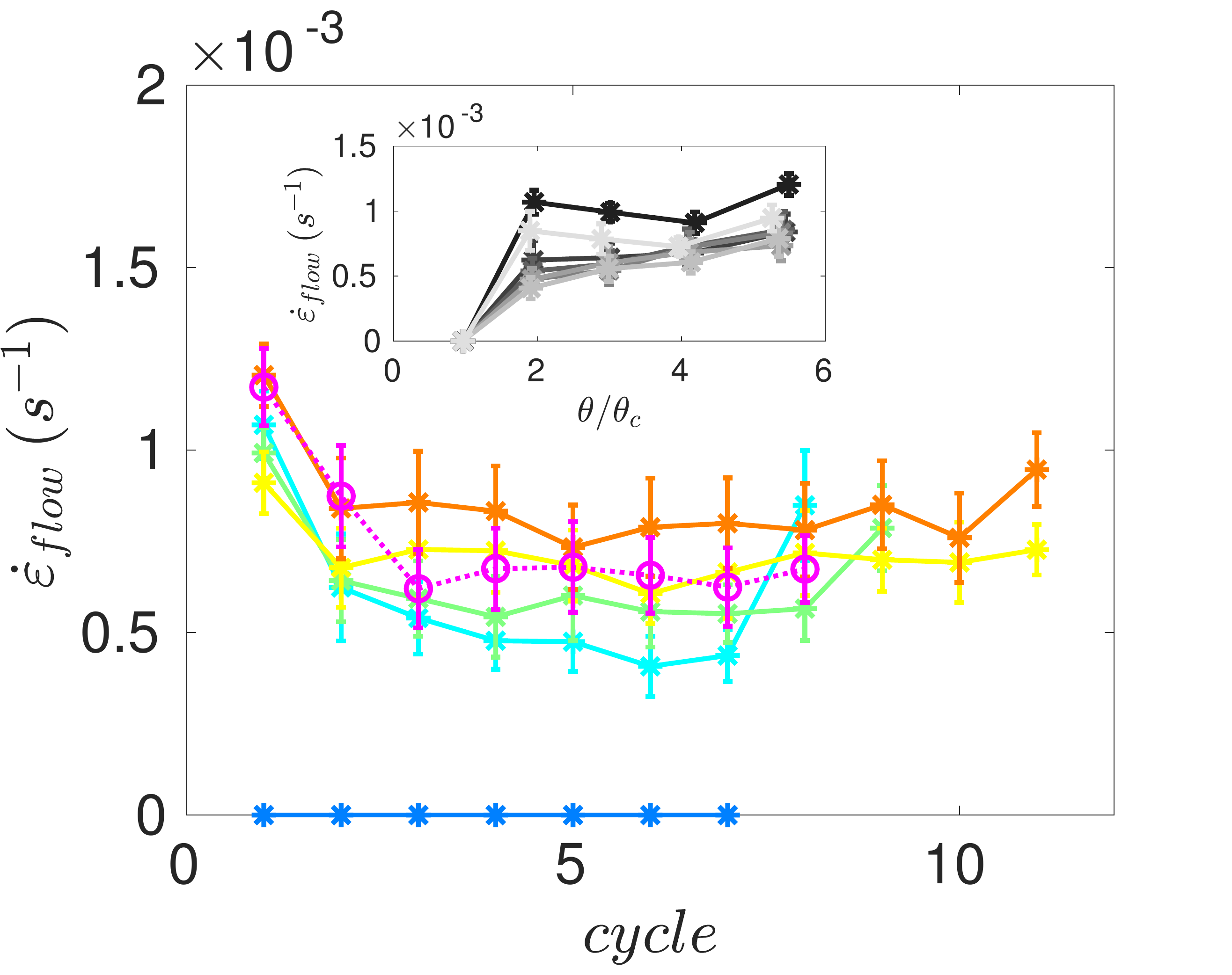}\\
		(b)
	\end{center}
\end{minipage}
\begin{minipage}[c]{0.32\linewidth}
	\begin{center}
		\includegraphics[width=\linewidth]{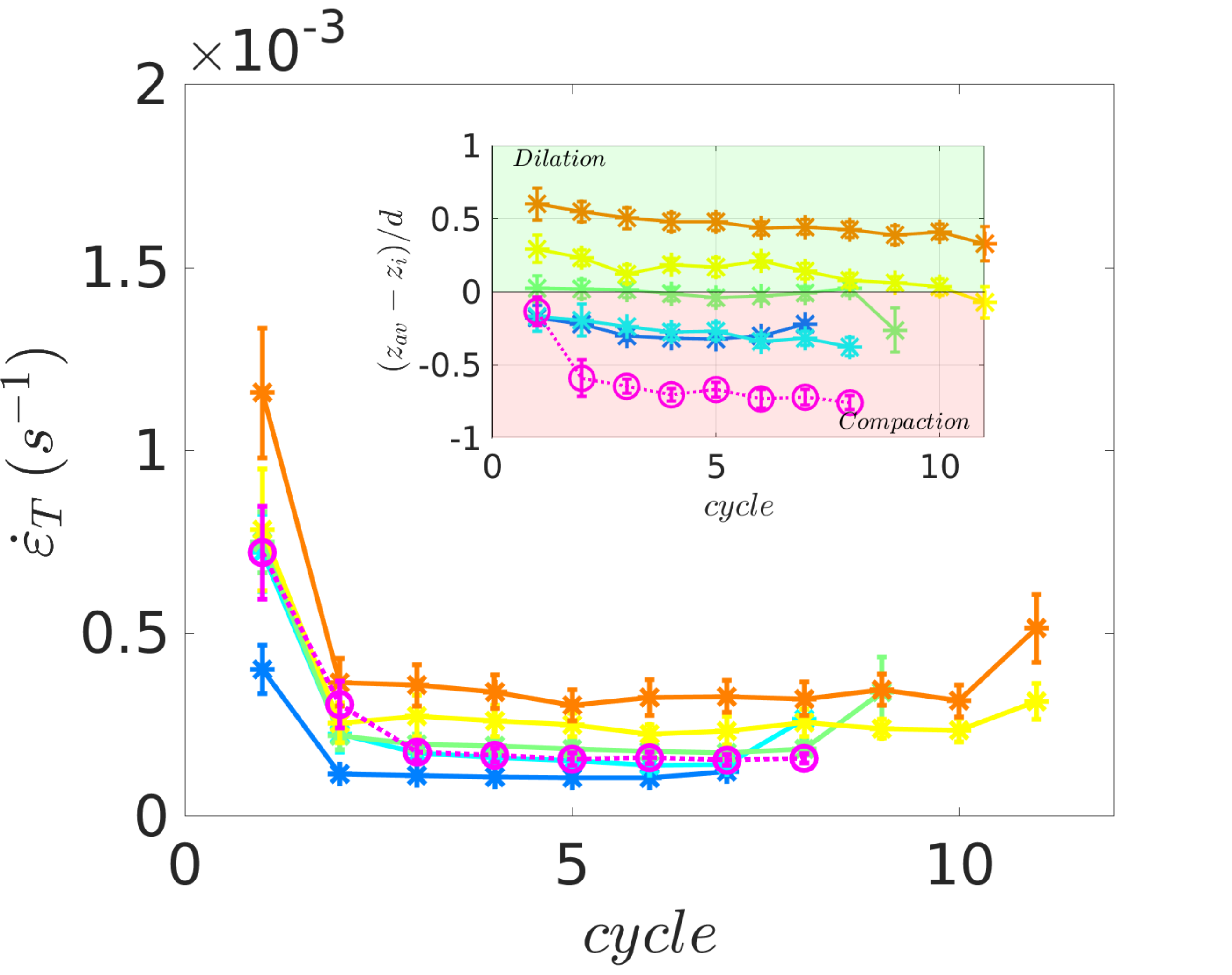}\\
		(c)
	\end{center}
\end{minipage}
\caption{Base and alternating cases. Strain rate measured in (a) the creeping part, (b) the flowing part, and (c) the total part of the sediment bed, as a function of time, or stress cycle; each cycle taking place over $\Delta t= 600 s$. Inserts: strain rate measured in (a) the creeping part and (b) the flowing part as a function of $\theta/\theta_c$, and (c) relative change of the bed surface elevation as a function of $\theta/\theta_c$. Gray levels scale with time, blue to orange colors represent low to high stresses. Curves of lightest level of gray correspond to the final cycle of base cases, when fluid shear is reverse. Pink curves correspond to the oscillatory scenario (also shown in figures 2b and 2d).  }
\label{fig_3}
\end{figure}

We first describe the phenomenology associated with unidirectional cyclic shear, followed by a direction reversal, in a typical run that we call the ``base case'' (Fig. \ref{fig_setup}). At time $t=0$ s a constant fluid shear stress of $\theta = 2\theta_{c}$ (i.e., $\theta = 0.2$) was applied in a clockwise direction. An unjamming front, initiated at the surface, propagated downward at progressively slower speed through time (Figs. \ref{fig_setup}(c) and \ref{displacements1}(a)). At $t = 10$ min the shear was turned off, and the time-integrated parameters were computed for the first cycle. Depth ($z$) profiles of average particle velocity and concentration followed the bi-partite patterns reported previously \cite{Houssais_1, houssais2016rheology}; $\langle V \rangle$ decreased rapidly with depth while $\langle C \rangle$ increased to a constant value in the bed-load layer, and below this layer $\langle V \rangle$ diminished more slowly (Fig. \ref{fig_setup}). The granular bed compacted by roughly 2 \% over the first stress cycle, as determined from a decrease in the bed-surface elevation (defined as the location at which $C/C_{max} = 0.5$ \cite{Houssais_1}). Repeated identical stress cycles showed progressive compaction and strain hardening of the granular bed --- as evidenced by reduced mobility of particles at all depths (Figs. \ref{displacements1} and \ref{fig_3}) --- though the effect of each subsequent cycle was diminished. Interestingly, the boundary between the upper bed-load layer and the lower creeping layer became sharper as the bed hardened, as indicated by the developing kink in the velocity profile. Using this kink to delineate the two regimes (cf. \cite{houssais2016rheology, Supplemental}), we can quantify changes in mobility in the creeping and bed-load layers separately (Fig. \ref{fig_3}). Strain rates for creep diminished more rapidly and significantly than bed load. Reversing the shear direction after seven cycles resulted in a (roughly) doubling of the strain rates for creep and bed load (Fig. \ref{fig_3}). Notably, this increased mobility was accompanied by a jump in compaction (Fig. \ref{fig_3}). 
This behavior is consistent with observations of shear jamming and fragile states in granular systems \cite{bi2011jamming}. In particular, while bed compaction and particle mobility appeared to saturate under a uni-directional stress, their abrupt increase on reversal indicates that the bed was conditioned to one direction but fragile to other directions. These results suggest that direction-dependent (anisotropic) strain hardening was smaller than, but of similar order to, (isotropic) compaction (Fig. \ref{fig_3}); and that memory of the former was erased by changing the shear direction. 

Repeating the base case experiment at higher $\theta$ values shows qualitatively similar behavior, but quantitative differences. As $\theta$ increases, the magnitude of strain hardening increases for creep but decreases for bed load (Fig. \ref{fig_3}). We attribute this to volume change and memory in the bed. Creep is generally associated with compaction, and the finding that increasing $\theta$ results in enhanced creep rates and faster strain hardening is in agreement with \citet{Allen_2}. Intense bed-load transport, however, is associated dilation and yielding, which erase memory \cite{Keim}. The response to shear direction reversal supports this interpretation; mobility is greatly enhanced for the creep regime for all $\theta$, but the jump in mobility for bed load diminishes as $\theta$ increases.

To further probe anisotropic effects, we compare a second ``alternating case'' experiment with $\theta/\theta_c = 2$ in which the only change from the base case is that shear direction was reversed for each cycle (Fig. \ref{fig_setup}). An important first observation is that bed compaction under alternating shear direction is much larger than all uni-directional experiments (Fig. \ref{fig_3}), consistent with previous work \cite{yang2021evolution}. If volume fraction were the only control on particle mobility, we would expect strain rates for the alternating case to be much lower than all uni-directional experiments. Particle mobility, however, remained elevated compared to the base case (Fig. \ref{fig_3}). We deduce that the alternating shear limited the development of a persistent granular fabric. Although shear reversal allowed the bed to find a more dense configuration, this did not make the bed stiffer compared to the uni-directional case. On the other hand, particle mobility in the alternating case was significantly lower than when shear was reversed at the end of the base case (Fig. \ref{fig_3}). We infer that the bed developed some kind of resisting fabric under bi-directional flow, similar to memory formation observed in oscillatory shear experiments \cite{Keim}; but that this fabric was not as strong as structures trained under uni-directional shear. 

\begin{figure}
\begin{minipage}[c]{0.49\linewidth}
	\begin{center}
		\includegraphics[width=.99\linewidth]{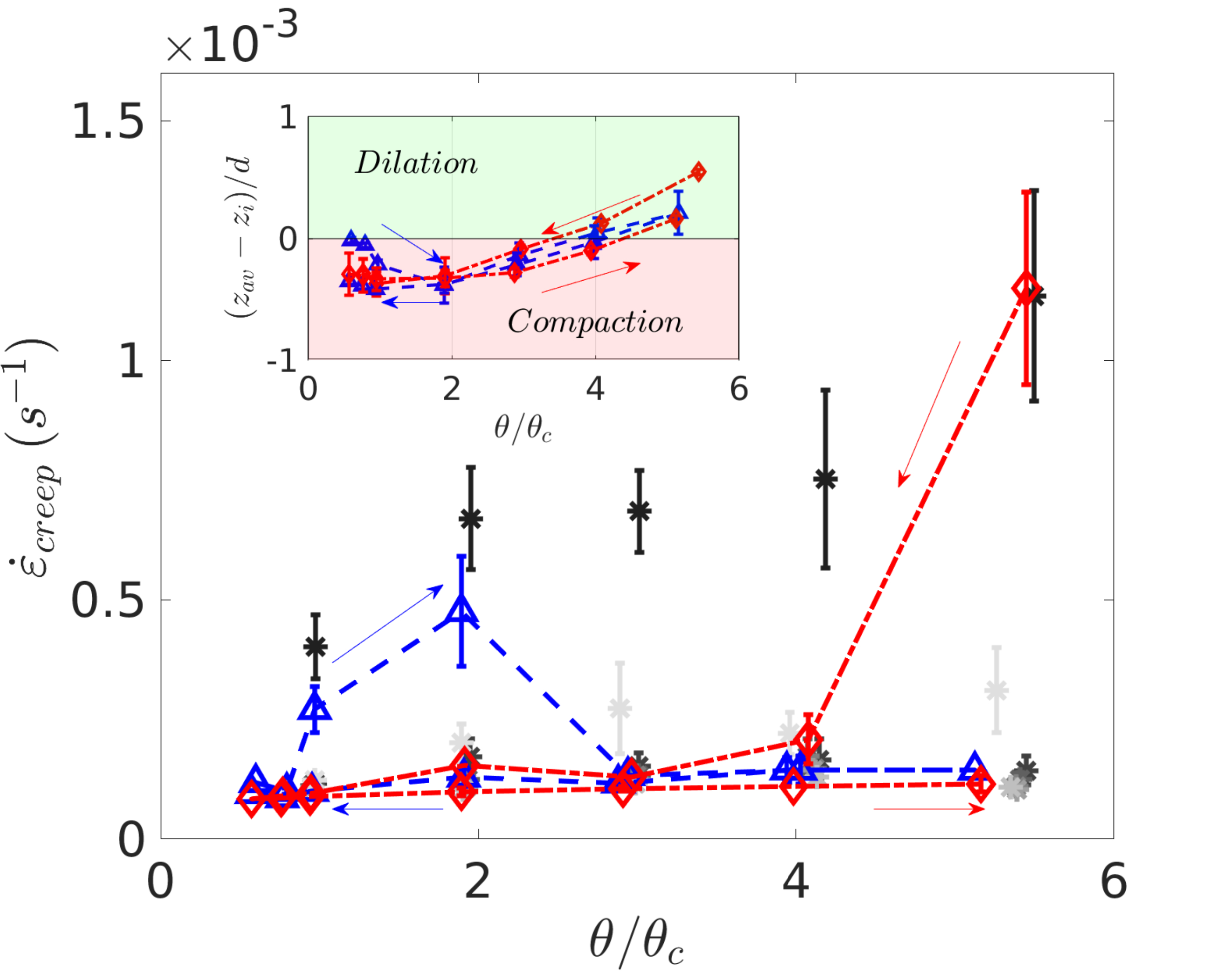}\\
		(a)
	\end{center}
\end{minipage}
\begin{minipage}[c]{0.49\linewidth}
	\begin{center}
		\includegraphics[width=.99\linewidth]{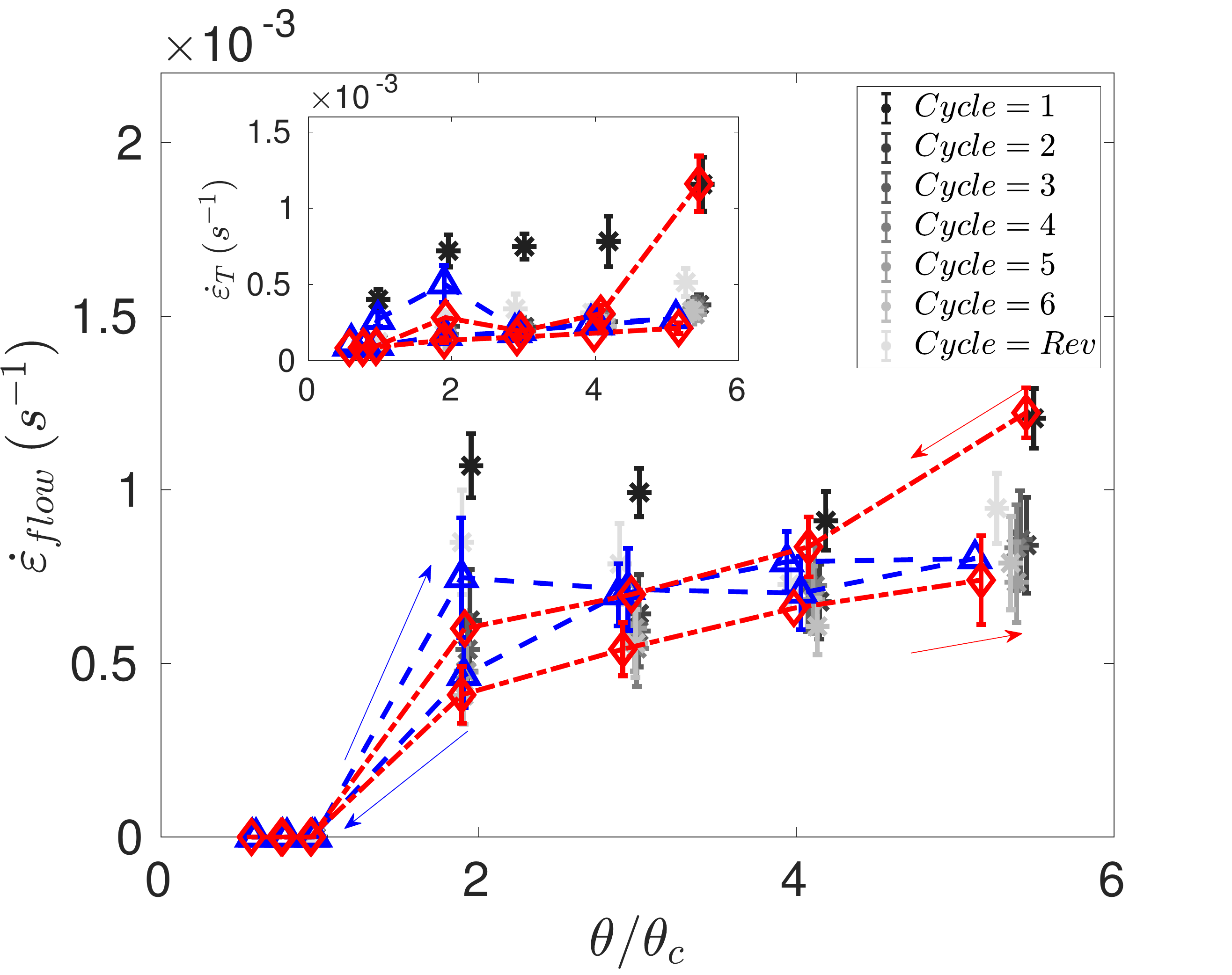}\\
		(b)
	\end{center}
\end{minipage}
\caption{Hysteresis cases. (a) Strain in the creeping part of the bed. Insert: relative change of the bed surface elevation as a function of $\theta/\theta_c$. (b) Strain  in the flowing part of the sediment bed, as a function of the dimensionless shear stress. Insert: total strain as a function of $\theta/\theta_c$. Gray levels scale with time, blue curves correspond to the increase-then-decrease stress scenario, red curves correspond to the decrease-then-increase stress scenario.  }
\label{fig_4}
\end{figure}

The above results reveal isotropic and anisotropic contributions to strain hardening, and that the latter may be erased by reversing flow. For uni-directional flows such as rivers, where shear reversal does not occur, these findings would lead us to expect only armored river beds. Clearly this is not the case; sufficiently large floods are known to break up armor and enhance sediment mobility \cite{vericat2006breakup}. It has been suggested that shear-induced dilation breaks up granular structures and weakens the bed \cite{Charru_1, Allen_2}, but this idea needs more systematic study. We performed two experiments to examine the effects of changing shear-stress magnitude under constant direction. Both experiments began with the same preparation protocol as above (i.e., an initially loose bed), and imposed stress sweeps with 13 cycles of 10 min each over a range (0.6-5)$\theta_c$ (Fig. \ref{fig_setup}). In the first sweep ($S \uparrow \downarrow$) stress was increased then decreased, and in the second sweep ($S \downarrow \uparrow$) stress was decreased and then increased. For both experiments we observed net compaction of the bed for $\theta < 4\theta_c$, and net dilation for larger stresses (Fig. \ref{fig_4}). We also confirmed that particle mobility decreased with compaction, and increased with dilation (Fig. \ref{fig_4}). We conclude, to first order, that dilation results from vigorous bed-load transport and acts to break up armoring, and compaction occurs under creep and weak bed load and forms armor.

The two stress sweeps, however, reveal a second-order history dependence. For $S \uparrow \downarrow$, the first stress cycle at 0.6$\theta_c$ initiated transient bed-load transport at the surface, that quickly died away. Small but measurable compaction occurred for sub-critical stress cycles as a consequence of creep, in agreement with \cite{Allen_2}. Persistent bed-load transport began at the surface for $\theta =  \theta_c$, and compaction increased with increasing stress up to 2$\theta_c$. Beyond this value, relative dilation occurred with increasing stress as the bed-load layer grew to consume more of the pack (Fig. \ref{fig_4}(a)). On the downward stress sweep, significant hysteresis was observed for $\theta <  2\theta_c$ because the strain-hardened and compacted bed could not return to its pre-sheared condition. These patterns were reversed for the $S \downarrow \uparrow$ sweep (Fig. \ref{fig_4}(b)). Starting from the highest stress the bed was more dilated relative to $S \uparrow \downarrow$, presumably because creep-induced compaction had not occurred. Relative compaction and declining mobility commenced with decreasing stress. Strong hysteresis was observed on the return (upward) stress sweep for $\theta >  2\theta_c$, where the strain-hardened and compacted bed suppressed dilation relative to the initial condition. These tests indicate that some memory of shear, in terms of strain hardening, exists even when applied stresses are $5\theta_c$. This memory is retained at depths that remain beneath the bed-load layer, and thus do not experience dilation; this implies that exceptionally large stresses capable of fluidizing the entire bed would be needed to erase all history of shear.

Compaction has long been known to cause stiffening of granular beds, due to increasing volume fraction that drives the system toward the jammed state \cite{richard2005slow}. The more recent recognition that materials may be driven towards jamming by shear, without any change in volume, has revealed the importance of anistropic grain fabric \cite{bi2011jamming, Cates, behringer2018physics}. Both of these factors are relevant in sediment transport, and our experiments found that they contribute comparably to strain hardening of the bed. Our findings add to recent experimental evidence \cite{Allen_2, deshpande2020perpetual} that creep is a primary driver of strain hardening and aging in granular beds. Creep may occur far below the entrainment threshold, and far beneath the bed-load layer. Interestingly, the separation between bed load and creep for an initially loose bed was not well defined (Fig. \ref{fig_setup}); the interface sharpened through time as a consequence of strain hardening. The emergence of a (more) mobile surface layer is reminiscent of granular heaps subject to tapping \cite{deshpande2020perpetual}, and also of the ``active layer'' description of mobile bed materials in natural rivers \cite{church2017active}. Due to the self-organization of river channels to a near-threshold state, fluid stresses rarely exceed $2\theta_c$ \cite{phillips2016self, phillips2019bankfull}. We thus expect that most flows strain harden the bed, while only exceptional floods break up surface armoring. This is in qualitative agreement with river observations that document changes in the threshold of motion as a function of flood history \cite{vericat2006breakup, turowski2011start, Masteller_2}. 
Further progress in comparing experiments to nature should take into account: the importance of collisions in water \cite{Pahtz_3, Pahtz_4, maurin2016dense}, which are damped in our viscous system; and also the shape and size dispersion of natural grains, which we expect to influence the rate but not the style of creep and bed load \cite{komatsu2001creep, ferdowsi2017river, gomez1994effects}.

F.D.C is grateful to FAPESP (Grants 2016/18189-0 and 2018/23838-3), E.M.F. to FAPESP (Grant 2018/14981-7), and P.A. and D.J.J. to ARO (Grant 579494) and NSF MRSEC (Grant 010401), for the financial support provided. We also thank B. Ferdowsi and C. Ortiz for initial help in running and analyzing experiments.

\bibliography{references}

\clearpage

\begin{center}
\textbf{SUPPLEMENTAL MATERIAL: Strain hardening by sediment transport}
\end{center} 

\section{Abstract}
This file contains supplementary information of two general types: additional detail on experimental techniques and analytical methods; and supplementary data figures that support the conclusions reported in the main text. We describe and illustrate how experimental quantities such as particle velocity, concentration, and strain were derived from raw images --- the principle data product of our study. We then present results of strain hardening and hysteresis for a variety of experiments, showing how our findings are qualitatively similar to those described in the main text under a range of conditions. Finally, this file includes a table showing all experimental conditions explored in the study.

\section{Determining particle velocity and volume fraction}
\begin{figure} [h!]
	\begin{minipage}{0.6\textwidth}
		\begin{tabular}{c}
			\includegraphics[width=\linewidth]{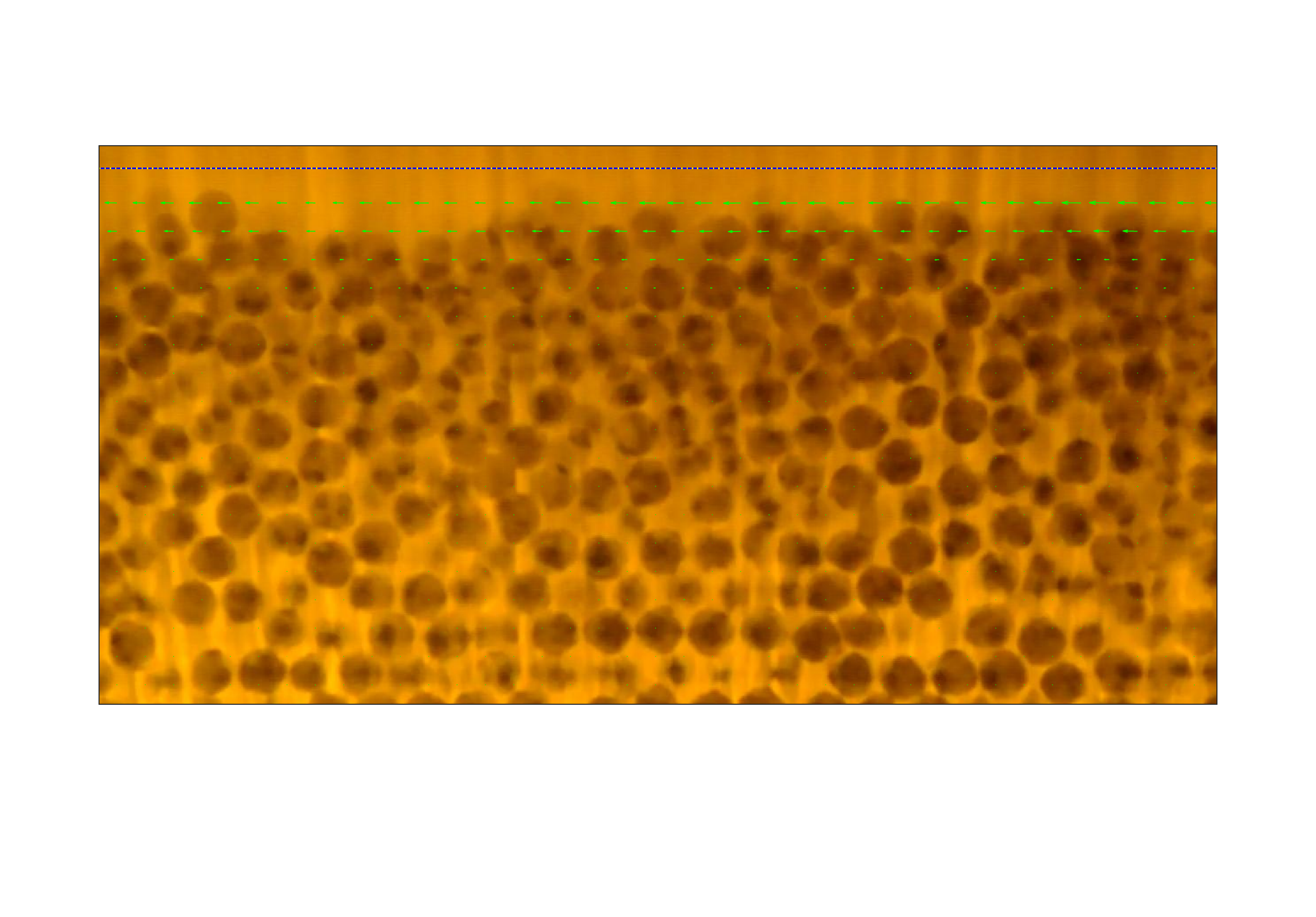}\\
			(a)
		\end{tabular}
	\end{minipage}
	\hspace{\fill}
	\begin{minipage}{0.5\textwidth}
		\centering
		\begin{tabular}{c}
			\includegraphics[width=\linewidth]{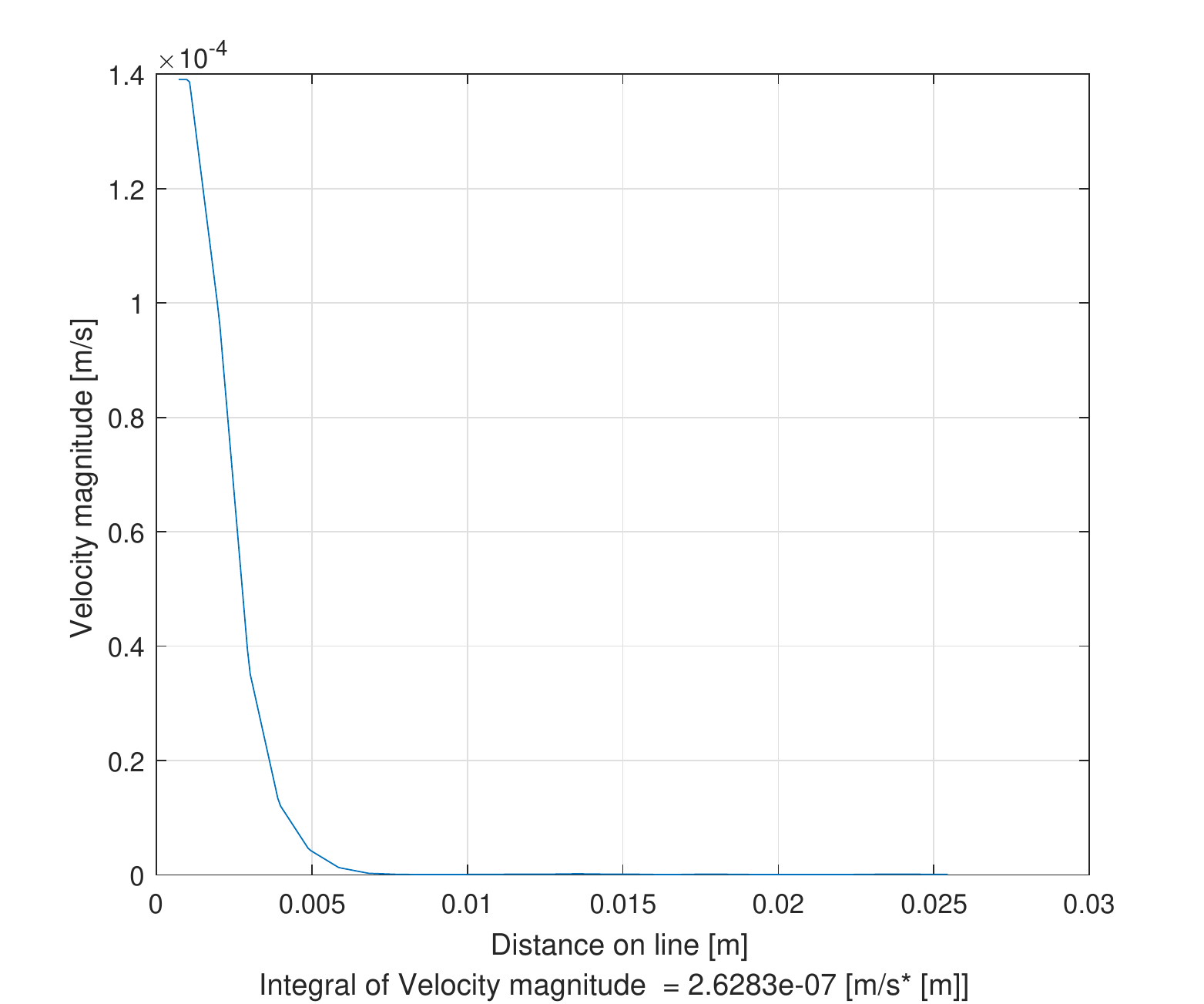}\\
			(b)
		\end{tabular}
	\end{minipage}
	\caption{Cross-correlation of sequential images to determine velocity, using PIV-LAB: (a) Example of ``instantaneous'' velocities obtained from PIV-LAB; and (b) averaged velocity profile.}
	\label{PIV}
\end{figure}

We examined a region of the sediment bed that is 1920 px in the horizontal (\textit{x}) direction and 960 px in the vertical ($z$) direction, corresponding to 48.5 mm $\times$ 24.23 mm. Images were obtained at the channel centerline (Fig. \ref{PIV} (a)) at a rate of 60 Hz. We first computed spatially-resolved (two-dimensional, 2D) ``instantaneous'' velocity measurements from successive image pairs by performing Particle Image Velocimetry (PIV) using PIV-LAB [1,2], with an interrogation area of 64 px $\times$ 64 px and an overlap of 50$\%$. This corresponds to 60$^2$ interrogation areas, and a spatial resolution of 0.8 mm $\times$ 0.8 mm.  Figure \ref{PIV} shows an example of the obtained velocities for the region of interest. Average profiles of velocity with depth, determined for each stress cycle, were then computed by $x-$averaging all instantaneous velocity measurements and then time-averaging all $x-averaged$ values over the 10-min duration of a stress cycle (Fig. \ref{PIV}). 

To estimate the particle volume fraction in 2D, we first detected all particles in each image using the method presented by Houssais et al. [3]. Profiles of 2D volume (area) fraction for each depth $z$ were then computed from the area occupied by the detected particles in each $x$ strip, averaged in time over the duration of each stress cycle.  

\section{Determining strain and strain rate}
\begin{figure}
	\includegraphics[width=0.9\linewidth]{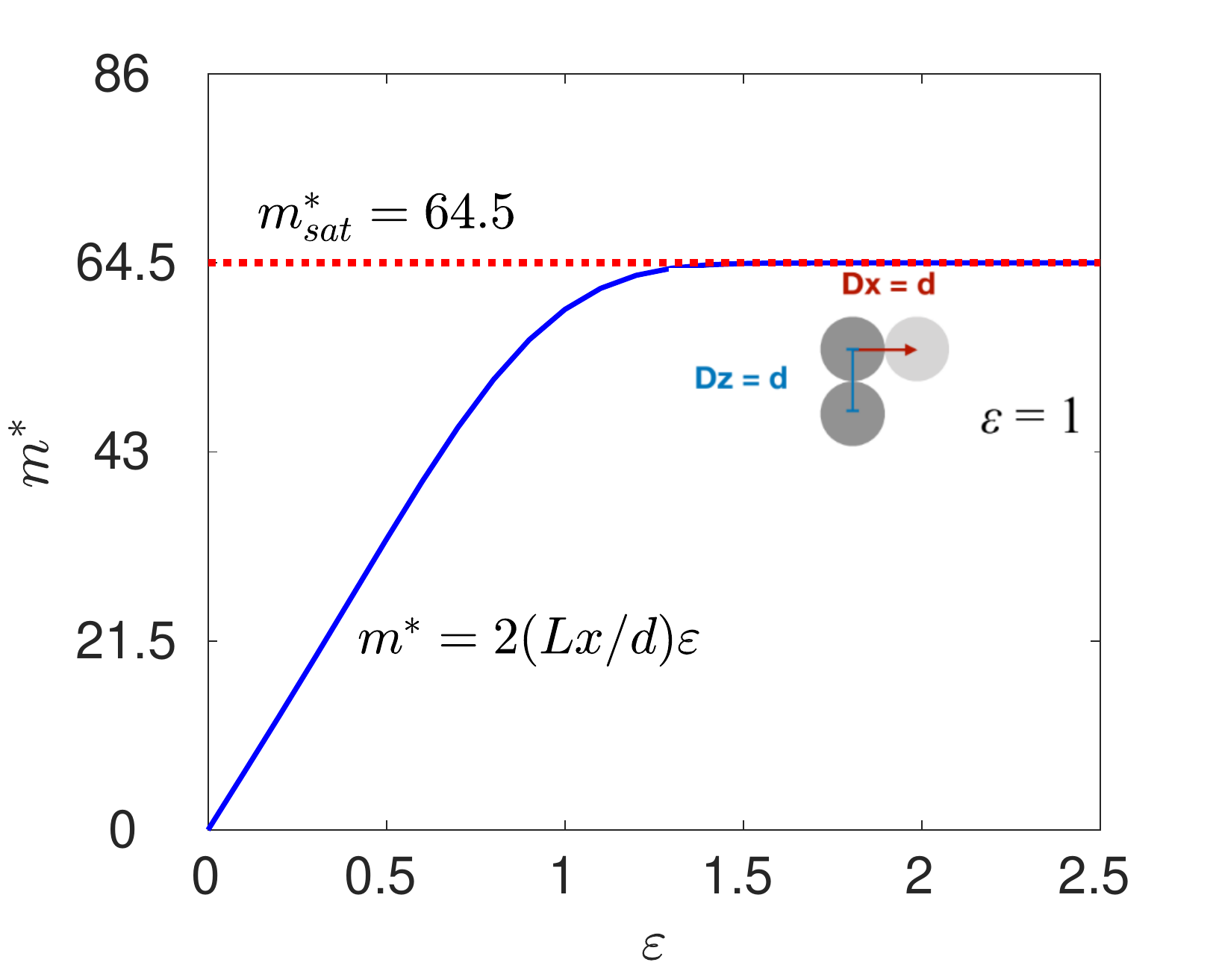}
	\caption{Dimensionless mobility as a function of strain, showing saturation at a strain of 1 that defines our measurement limit.}
	\label{ms}
\end{figure}

Strain was determined for each pixel in the area of interest of a fixed duration of the stress cycle, $\Delta t = 600 s$. We examined the change in pixel intensity inside image strips of width $L_x$ and height $d$, a quantity we call mobility $m$ [$px^2$]. If one particle of diameter $d$ moves laterally in that strip over a distance $\Delta x = d$ -- which corresponds to the case where the strain $\varepsilon = \Delta x/ \Delta z = 1$ -- then $m =2 \pi (d/2)^2$ [$px^2$]. In the case of $n$ particles in a dense configuration that move along a strip of lateral size $L_x$, it should be the case that $L_x = n*d$. We now define the dimensionless parameter $m^* = m\: (scale)^2/\pi (d/2)^2$, where $scale$ is the resolution [m/pxl] in the image. For the case of one particle exhibiting $\epsilon = 1$, $m^* = 2$. For the case of $n$ particles in the dense configuration described above, $m^* = 2 n = 2 L_x/d$. This case is the largest deformation that can be monitored by measuring pxl change; when $\varepsilon > 1$, some particle displacement is not captured. We therefore determine a saturation value of the mobility measurement as $m^*_{sat} = 2 L_x/d$; for our images, $m^*_{sat} = 64.5$ (Fig. \ref{ms}). 

Using the time-averaged concentration profile $\left<C(z)\right>(z)$, we define the local and time-averaged measurement of strain, over a duration $\Delta t = 600 s$, and for $m^*<m^*_{sat}$, as:
\begin{equation}
	\left<\varepsilon(z)\right>=\frac{C_{sat}}{\left<C(z)\right>}\frac{d}{2Lx}m^*(z) 
\end{equation}
\noindent where $m$ [$px^2$] comes from the absolute difference between two images. To obtain the $m$ profiles at each time step, we compared the first image of each cycle with the next ones until $t=600 s$, i.e., $m(t)=abs(image(1)-image(t))$.

The average strain value in the creep regime is given by the expression:
\begin{equation}
	\varepsilon_{creep}=\frac{1}{z_{c}}\int_{0}^{z_c}\left<\varepsilon(z)\right>dz
\end{equation}
\noindent where $z_c$ is the depth associated with the kink in the mean velocity profile, that separates the creep and dense-flow regimes [3]. The average strain value in the flowing regime is given by:

\begin{equation}
	\varepsilon_{flow}=\frac{1}{z|_{m^*=m^*_{sat}}-z_{c}}\int_{z_c}^{z|_{m^*=m^*_{sat}}}\left<\varepsilon(z)\right>dz
\end{equation}
\noindent where $z|_{m^*=m^*_{sat}}$ is the depth of the bed where $m^*$ reaches $m^*_{sat}$. Finally, the total average strain follows the expression:
\begin{equation}
	\varepsilon_{T}=\frac{1}{z|_{m^*=m^*_{sat}}}\int_{0}^{z|_{m^*=m^*_{sat}}}\left<\varepsilon(z)\right>dz
\end{equation} 
\clearpage


\textbf{References} 

[1] W. Thielicke and E. Stamhuis, Pivlab–towards user-friendly, affordable and accurate digital particle image velocimetry in matlab, Journal of open research software 2 (2014).
 
[2] W. Thielicke, The flapping flight of birds, Diss. University of Groningen (2014).

[3] M. Houssais, C. P. Ortiz, D. J. Durian, and D. J. Jerolmack, Onset of sediment transport is a continuous transition driven by fluid shear and granular creep, Nat. Commun. 6 (2015).

\section{Supplementary Figures}

\begin{figure}[h!]
	\centering
	\includegraphics[angle=0, scale=0.6, clip]{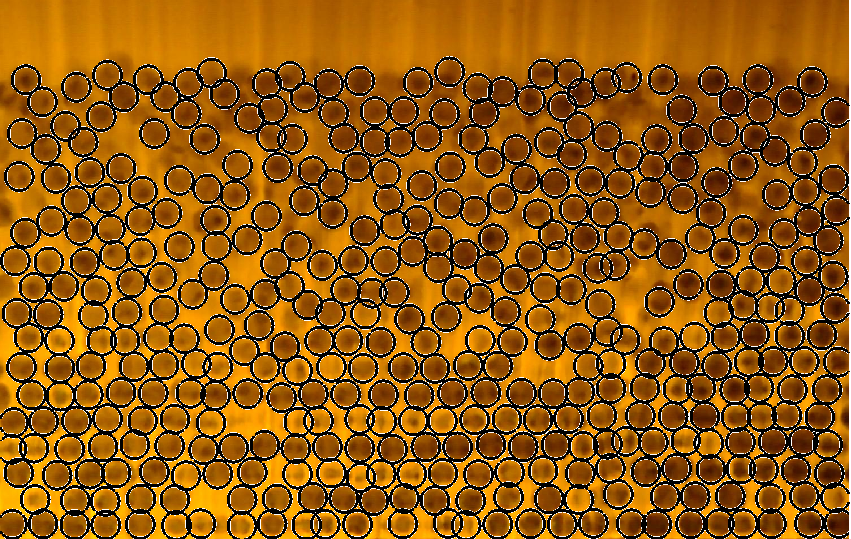}
	\caption{Example image showing detection of particles, using the method described in [3]. 
	}
	\label{detec}
\end{figure}   

\begin{figure}
	\includegraphics[width=1\linewidth]{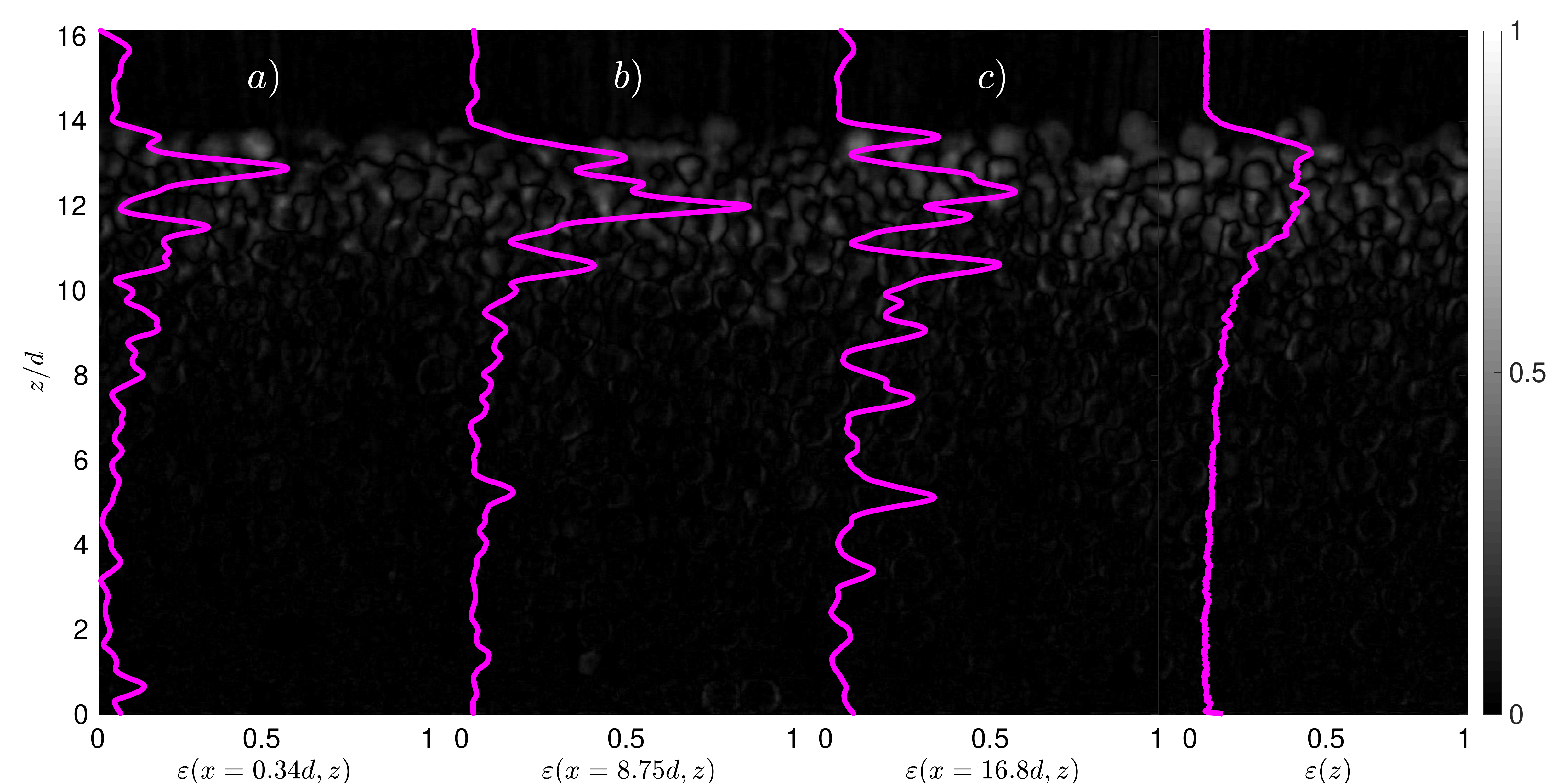}
	\caption{Local and spatially-averaged strain measurements. Background shows local strain measurements for each pixel.		Profiles are superimposed for strain as a function of depth, $\varepsilon (z)$, at different locations $x$ in the image, for an experiment with $\theta/\theta_c=2$ at the last stress cycle and at time $t=600 s$. Strain profiles at (a) location $x=0.34d$, (b) location $x=8.75d$, (c) location $x=16.8d$, and (d) averaged spatially across all locations $x$. }
	\label{fig2}
\end{figure}

\begin{figure}
	\includegraphics[width=1\linewidth]{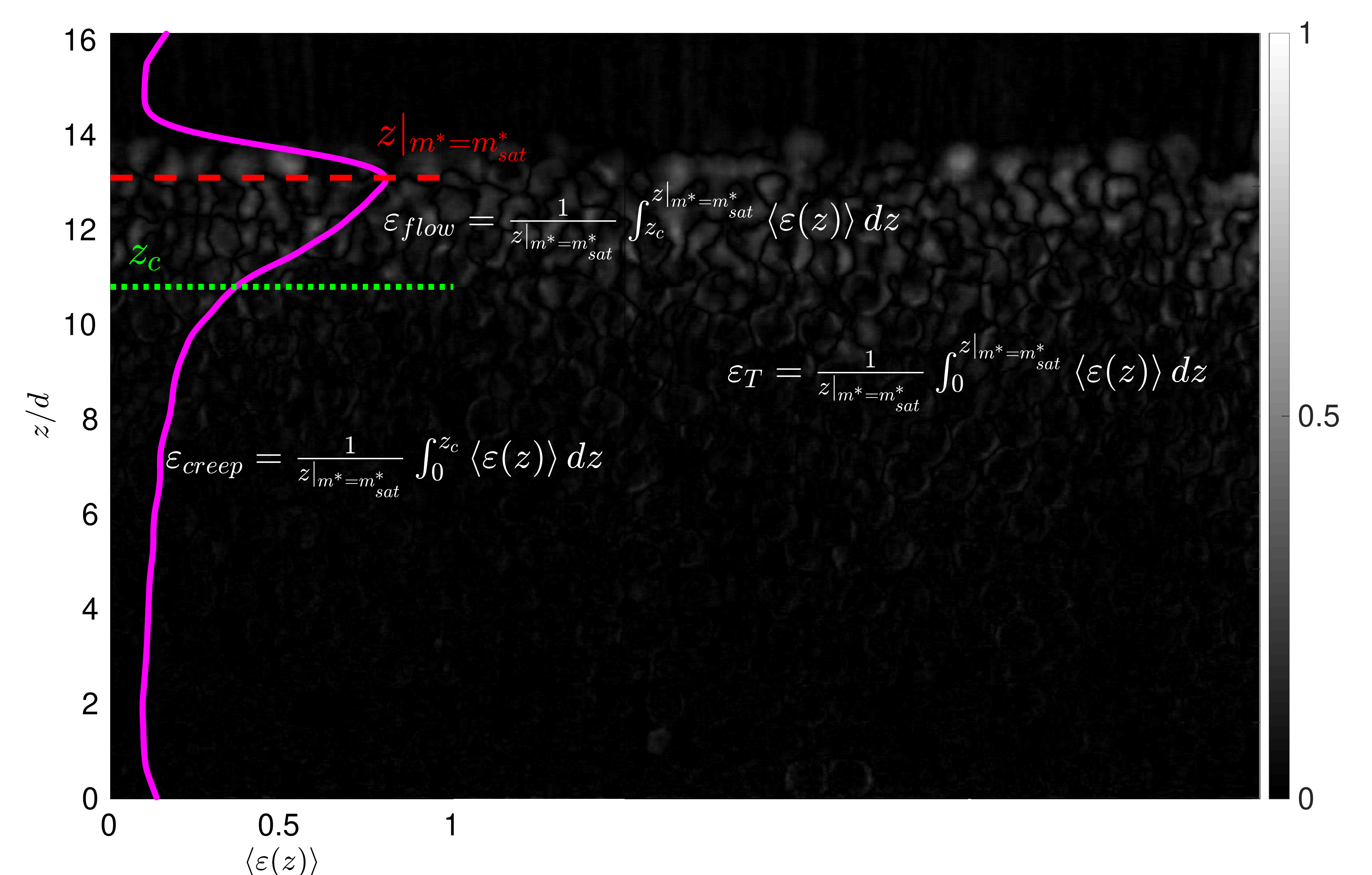}
	\caption{Local and time-averaged strain for different regimes for $\theta/\theta_c=2$, computed over the last stress cycle.		Equations used to determine strain for Creep, Flowing and Total (Creep + Flowing) are shown.}
	\label{fig3}
\end{figure}

\begin{figure}[ht]
	\begin{minipage}{0.49\linewidth}
		\begin{tabular}{c}
			\includegraphics[width=1\linewidth]{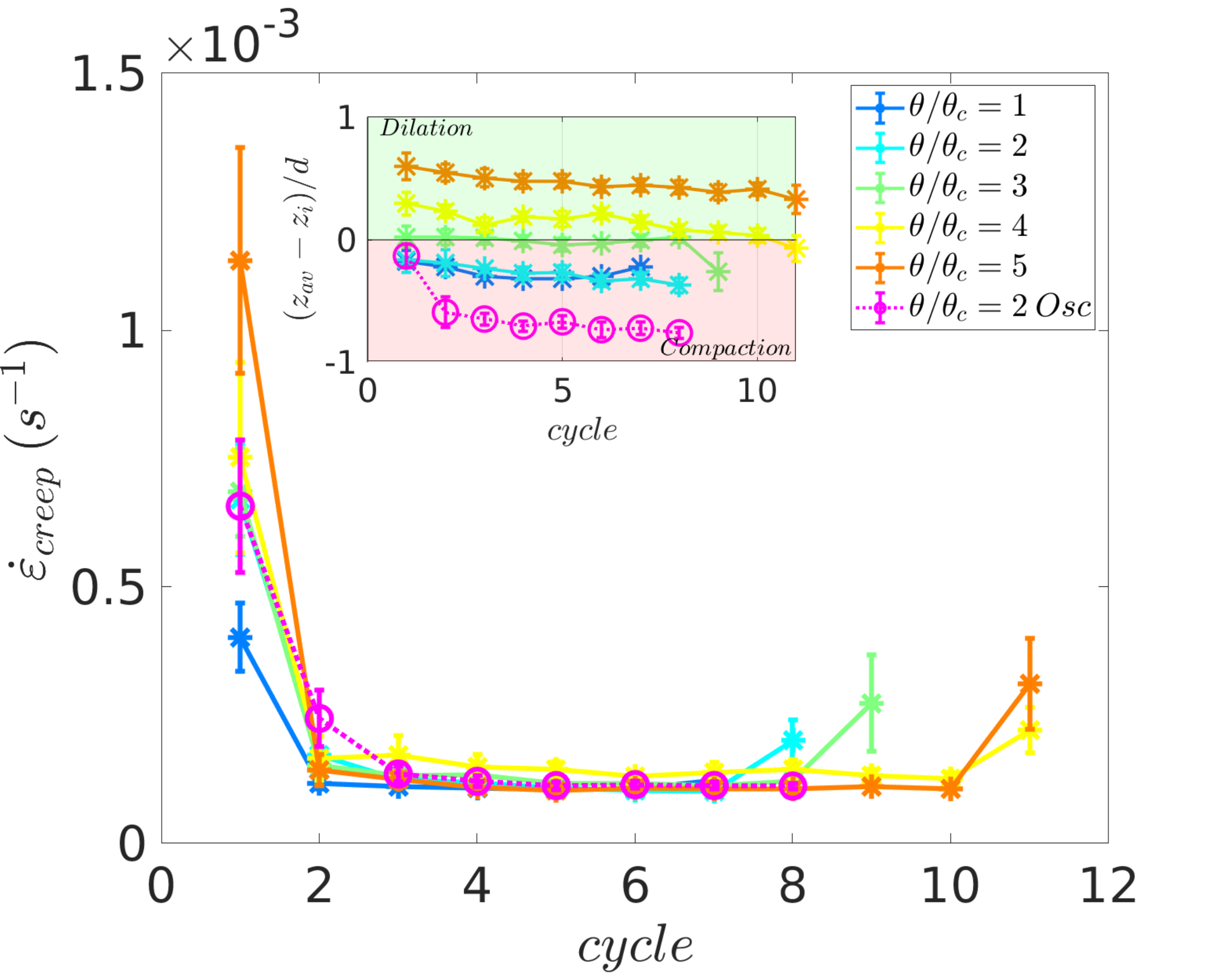}\\
			(a)
		\end{tabular}
	\end{minipage}
	\hfill
	\begin{minipage}{0.49\linewidth}
		\begin{tabular}{c}
			\includegraphics[width=1\linewidth]{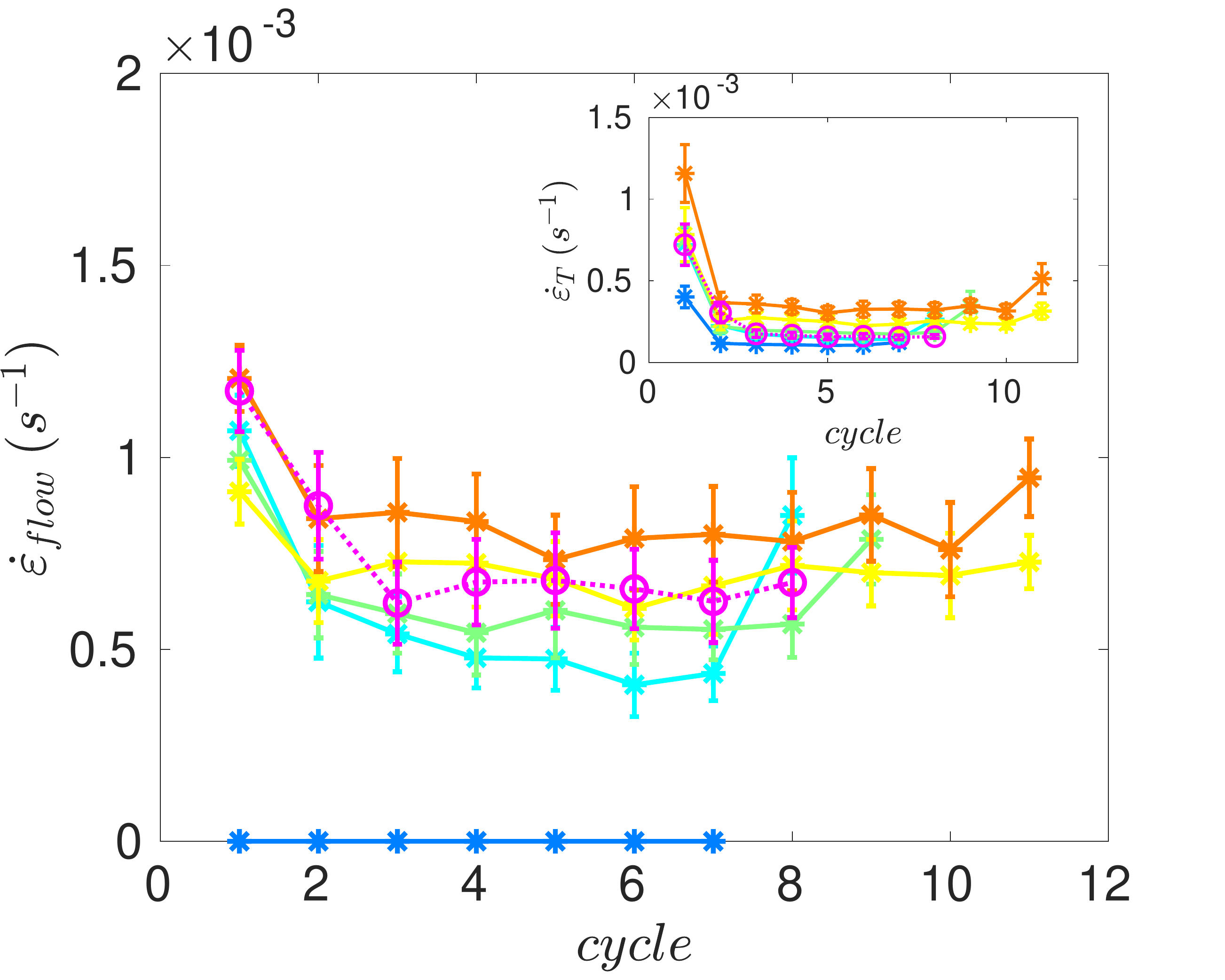}\\
			(b)
		\end{tabular}
	\end{minipage}
	\caption{(a) Strain rate in the creeping regime as a function of the number of stress cycles, for uni-directional experiments where shear direction was reversed only for the last cycle; inset shows the height of the sediment bed, determined as described in the main text. (b) Strain rate in the flowing regime as a function of the number of cycles; inset shows the total strain rate. Data presented for tests 1-5 and 8 in Table \ref{tabex1}); legend in (a) indicates associated stress.}
	\label{fig:strain1}
\end{figure}  

\begin{figure}[ht]
	\begin{minipage}{0.49\linewidth}
		\begin{tabular}{c}
			\includegraphics[width=1\linewidth]{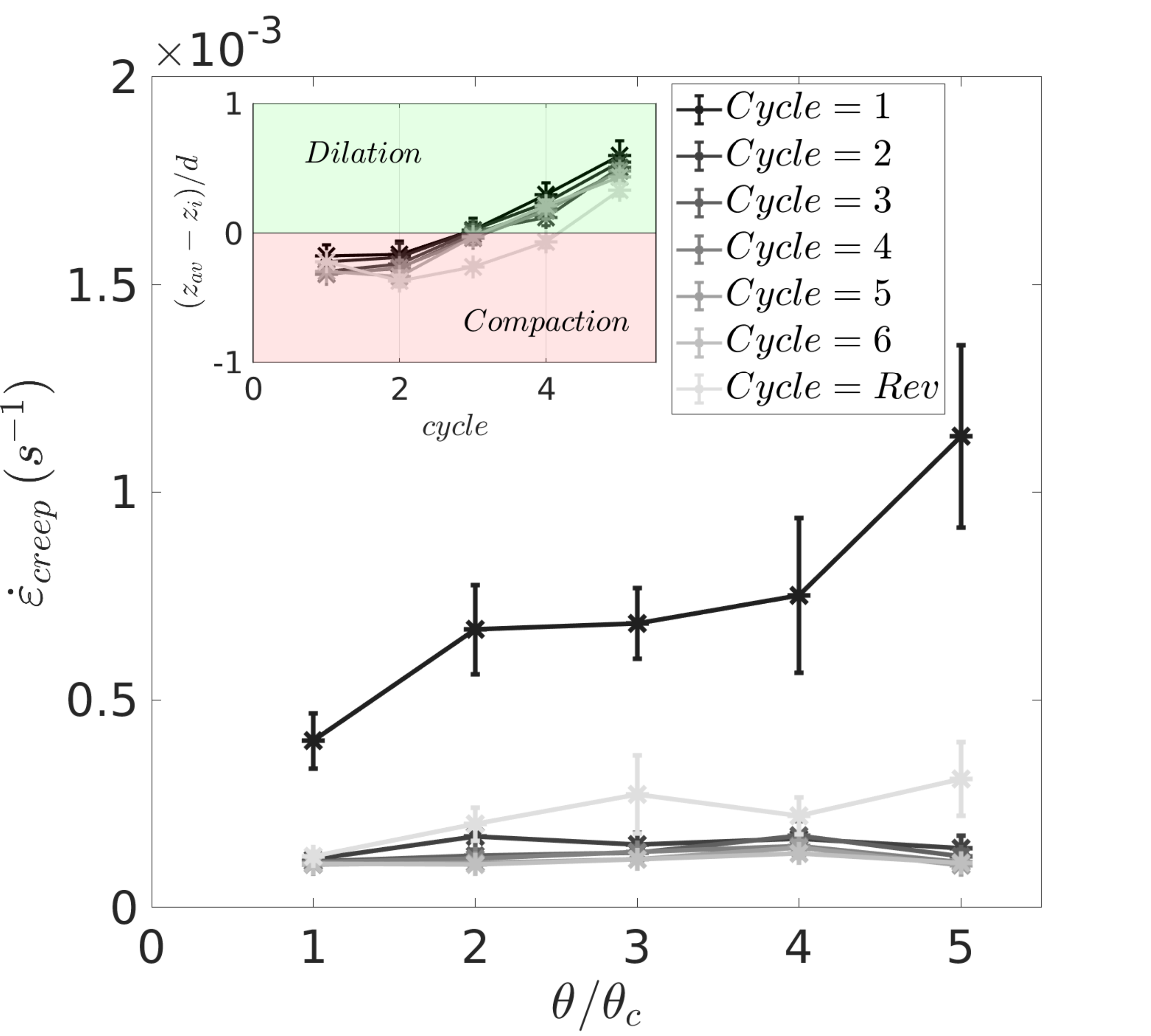}\\
			(a)
		\end{tabular}
	\end{minipage}
	\hfill
	\begin{minipage}{0.49\linewidth}
		\begin{tabular}{c}
			\includegraphics[width=1\linewidth]{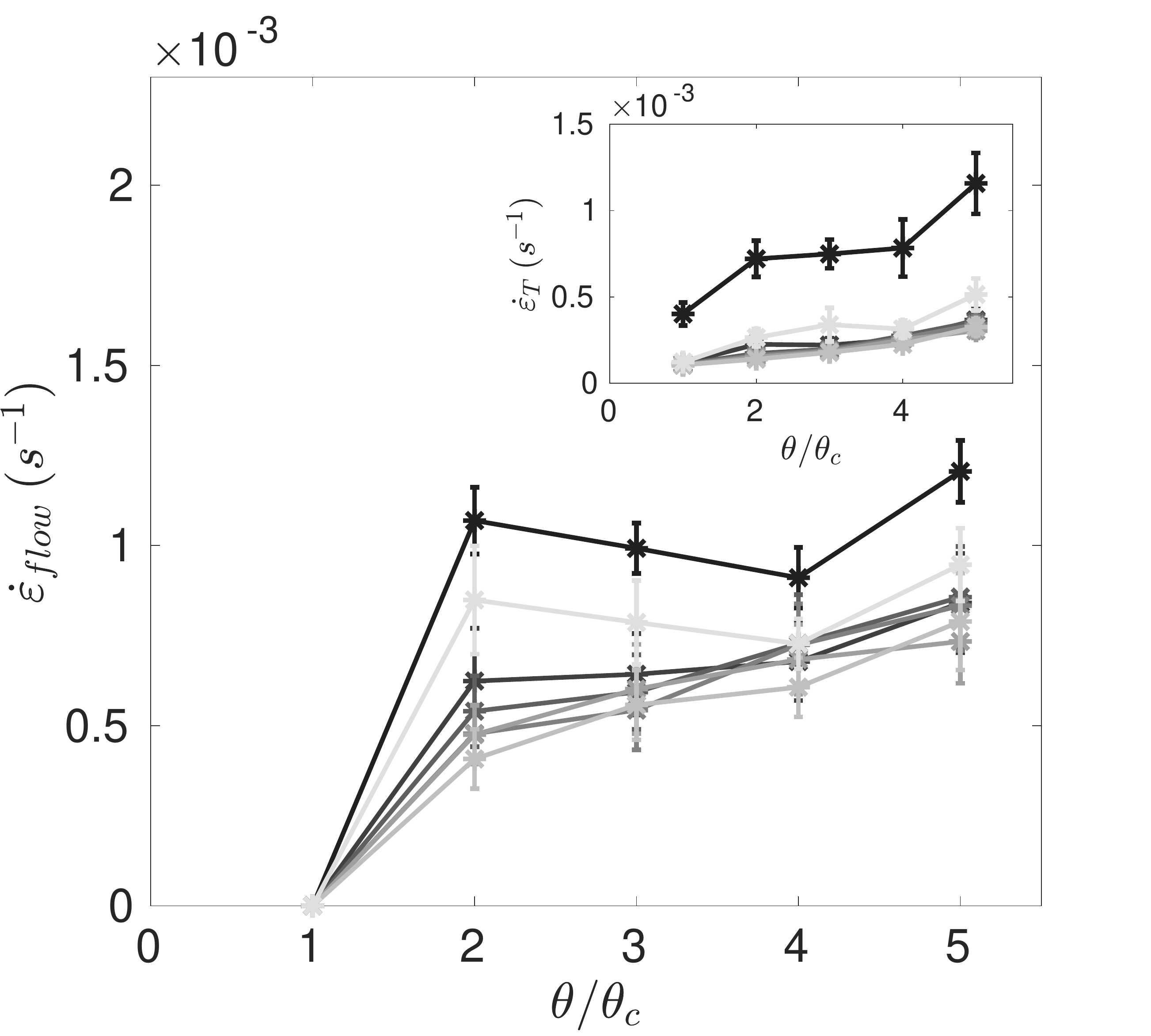}\\
			(b)
		\end{tabular}
	\end{minipage}
	\caption{An alternative way to present the data in Figure \ref{fig:strain1}; strain rate as a function of the applied shear stress, where grayscale corresponds to the stress cycle. (a) Strain rate for the creeping regime; and (b) strain rate for the flowing regime. Insets follow Figure \ref{fig:strain1}; results shown for tests 1-5 in Tab.\ref{tabex1}.}
	\label{fig:straint1}
\end{figure}

\begin{figure}[b]
	\includegraphics[width=0.9\linewidth]{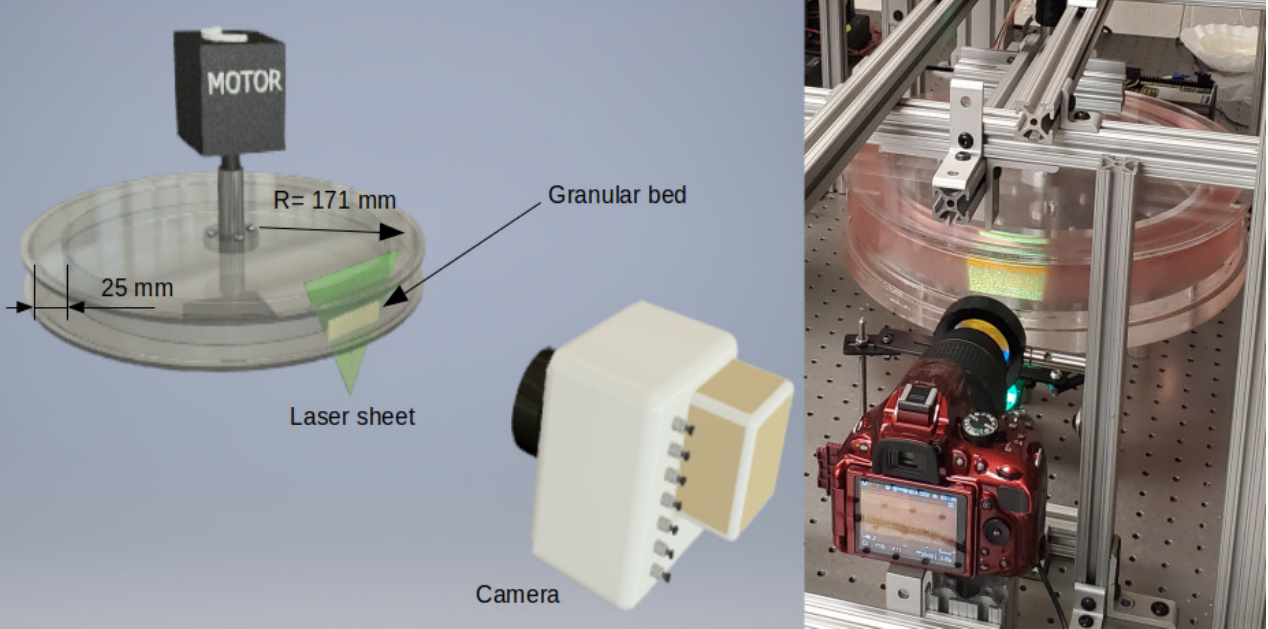}
	\caption{A layout (on the left) and a photograph (on the right) of the experimental setup.}
	\label{fig_setup}
\end{figure}

\begin{figure}[b]
	\includegraphics[width=0.8\linewidth]{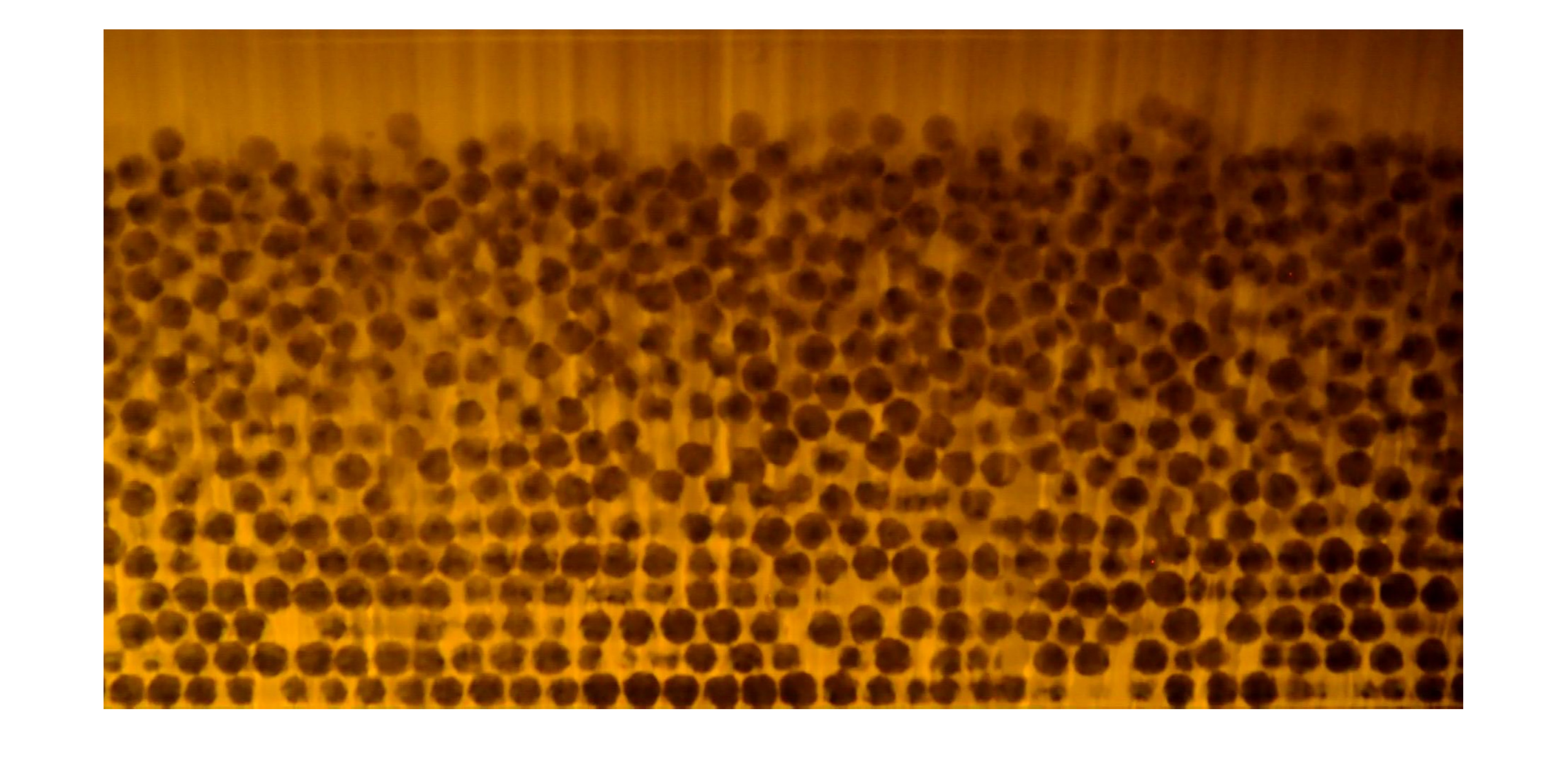}
	\caption{Raw image of the grains within the bed in the laser plane.}
	\label{raw_image}
\end{figure}

\begin{table}[h!t]
	\caption{Tested conditions. From left to right, the test number, Shields number normalized by its critical value, lid velocities at the channel centerline, number of cycles, shear direction during the preparation of the bed, shear direction during cycles, and direction of the last cycle. OBS: for test 6, $\theta/\theta_c=4$ in the last cycle.}
	\begin{center}
		\begin{tabular}{cccccccc}
			\hline
			Test & $\theta/\theta_c$ & $U_{lid}$ & number of cycles & preparation & cycles & last cycle\\
			$\cdots$ & $\cdots$ & $mm/s$ & $\cdots$ & $\cdots$ & $\cdots$ & $\cdots$\\
			\hline
			$1$ & $1$ & $9$ & $7$ & Same direction & Same direction & Reversed direction\\
			$2$ & $2$ & $19$ & $8$ & Same direction & Same direction & Reversed direction\\
			$3$ & $3$ & $28$ & $9$ & Same direction & Same direction & Reversed direction\\
			$4$ & $4$ & $37$ & $11$ & Same direction & Same direction & Reversed direction\\
			$5$ & $5$ & $46$ & $11$ & Same direction & Same direction & Reversed direction\\
			$6$ & $2$ & $19$ & $8$ & Same direction & Same direction & Same direction\\
			$7$ & $2$ & $19$ & $8$ & Reversed direction & Same direction & Reverse\\
			$8$ & $2$ & $19$ & $8$ & Same direction & Alternating direction & Reverse\\
			$9$ & $1-5-1$ & $9-46$ & $13$ & Same direction & Same direction & Same direction\\
			$10$ & $5-1-5$ & $46-9$ & $13$ & Same direction & Same direction & Same direction\\
			\hline
		\end{tabular}
		\label{tabex1}
	\end{center}
\end{table}  

\begin{figure}[b]
	\includegraphics[width=0.8\linewidth]{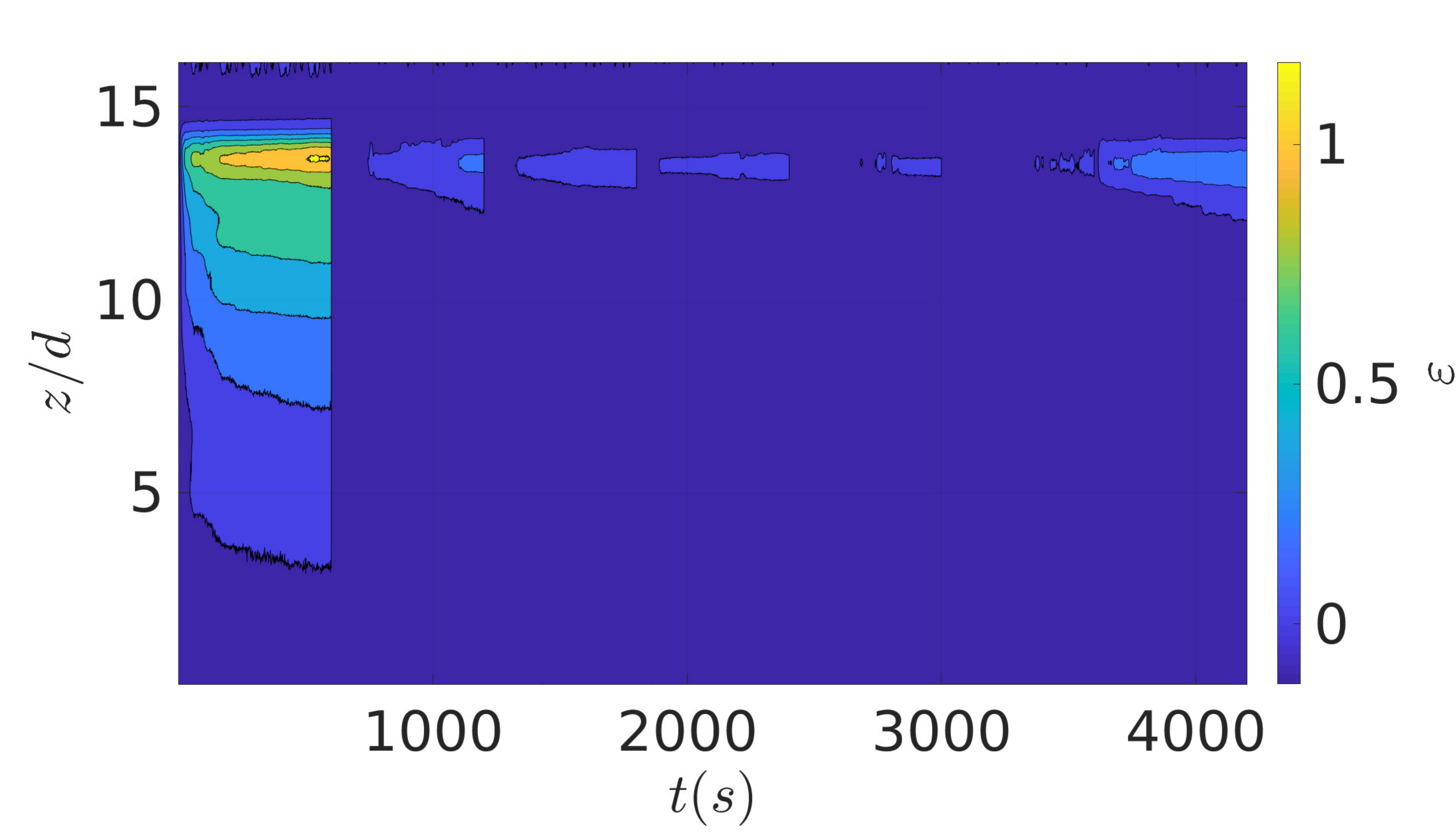}
	\caption{$M$ matrix representing local particles' mobility intensity for test 1 (all cycles included) with reversal at the last cycle. Variation from blue (darker shades in grayscale) to yellow (brighter shades in grayscale) corresponds to increasing in mobility.}
	\label{dif1}
\end{figure}

\begin{figure}[b]
	\includegraphics[width=0.8\linewidth]{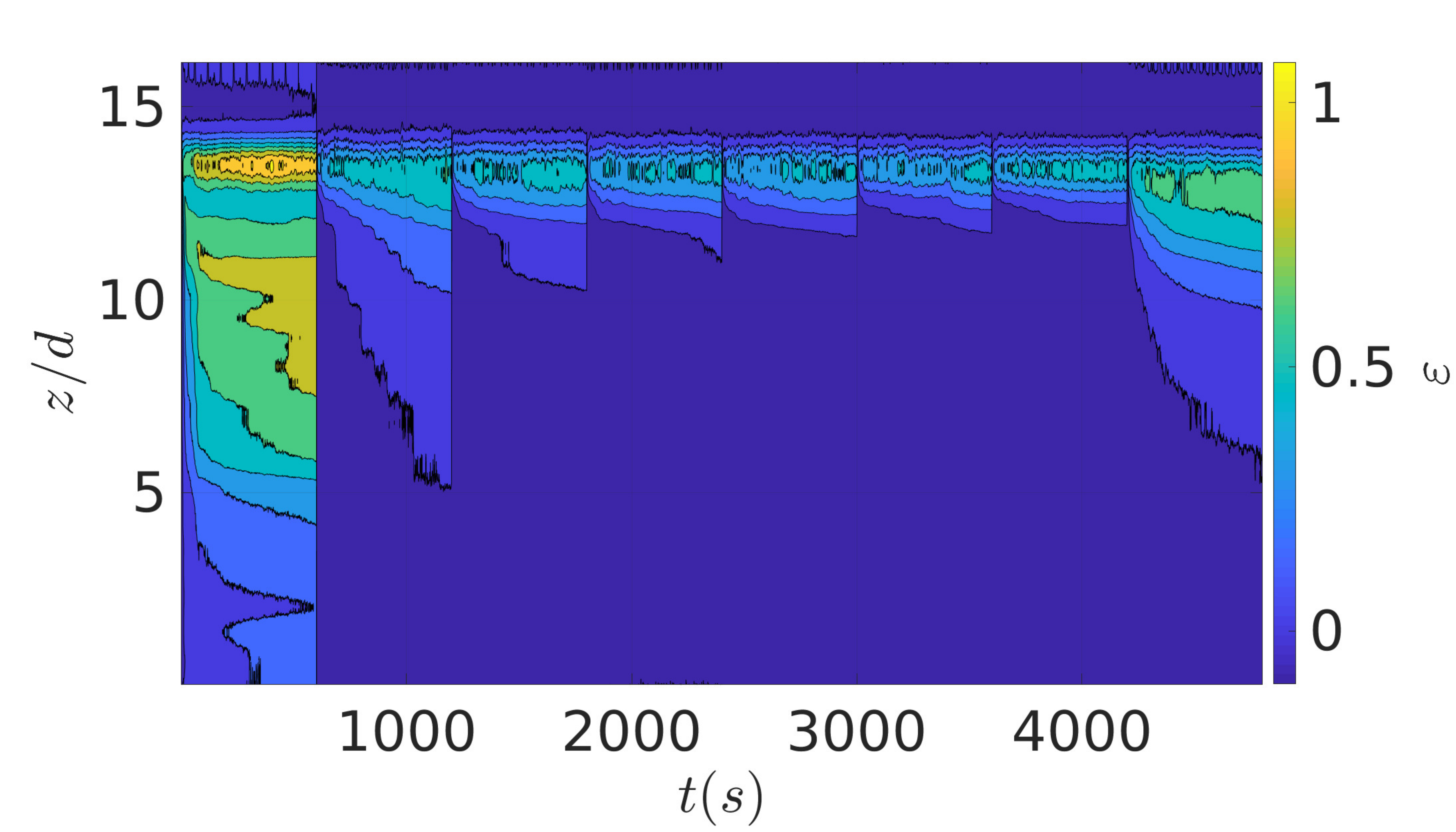}
	\caption{Same as Figure \ref{dif1}, but for test 2 (all cycles included).}
	\label{dif2}
\end{figure}

\begin{figure}[b]
	\includegraphics[width=0.8\linewidth]{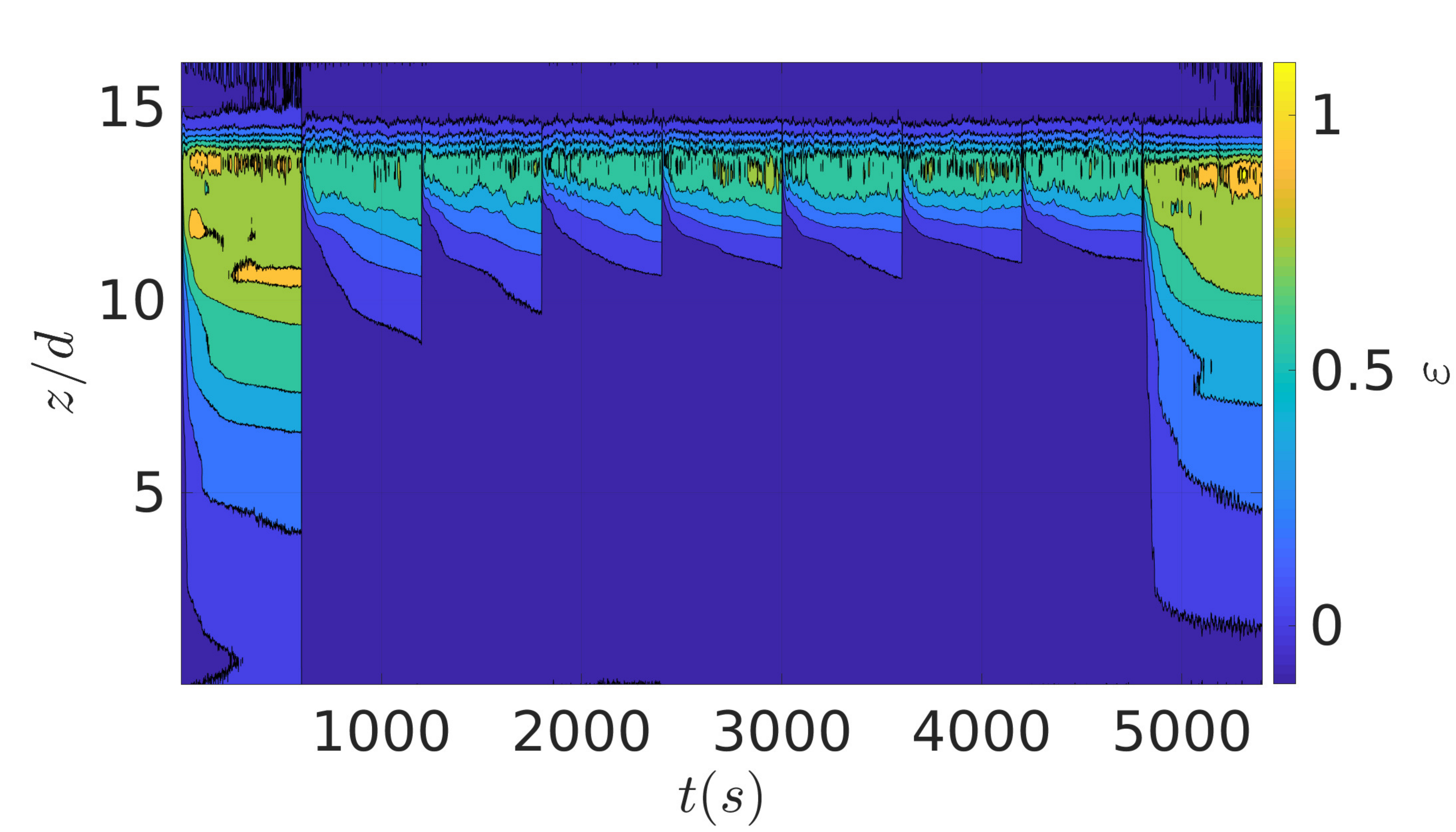}
	\caption{Same as Figure \ref{dif1}, but for test 3 (all cycles included)}
	\label{dif3}
\end{figure}

\begin{figure}[b]
	\includegraphics[width=0.8\linewidth]{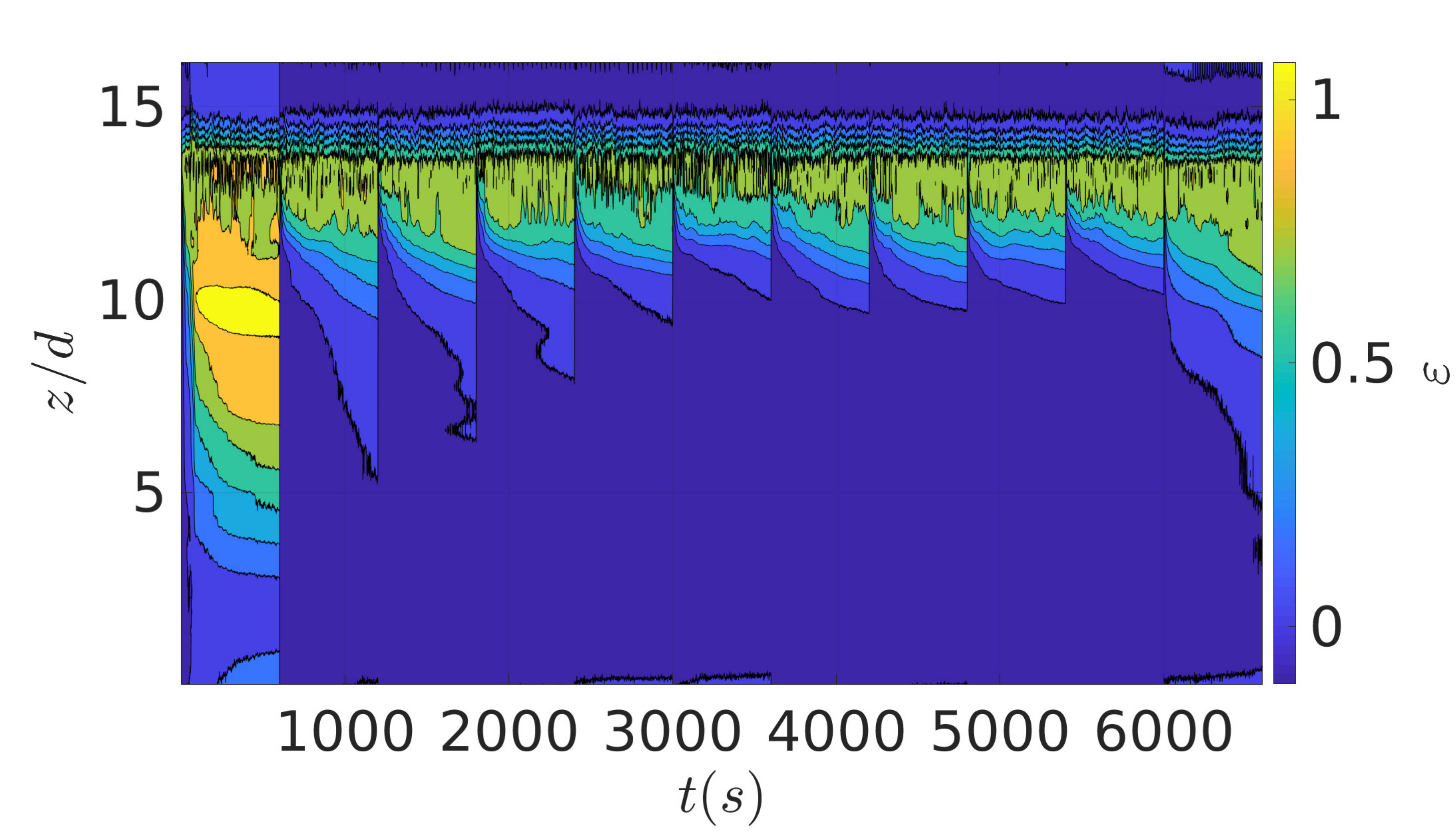}
	\caption{Same as Figure \ref{dif1}, but for test 4 (all cycles included)}
	\label{dif4}
\end{figure}

\begin{figure}[b]
	\includegraphics[width=0.8\linewidth]{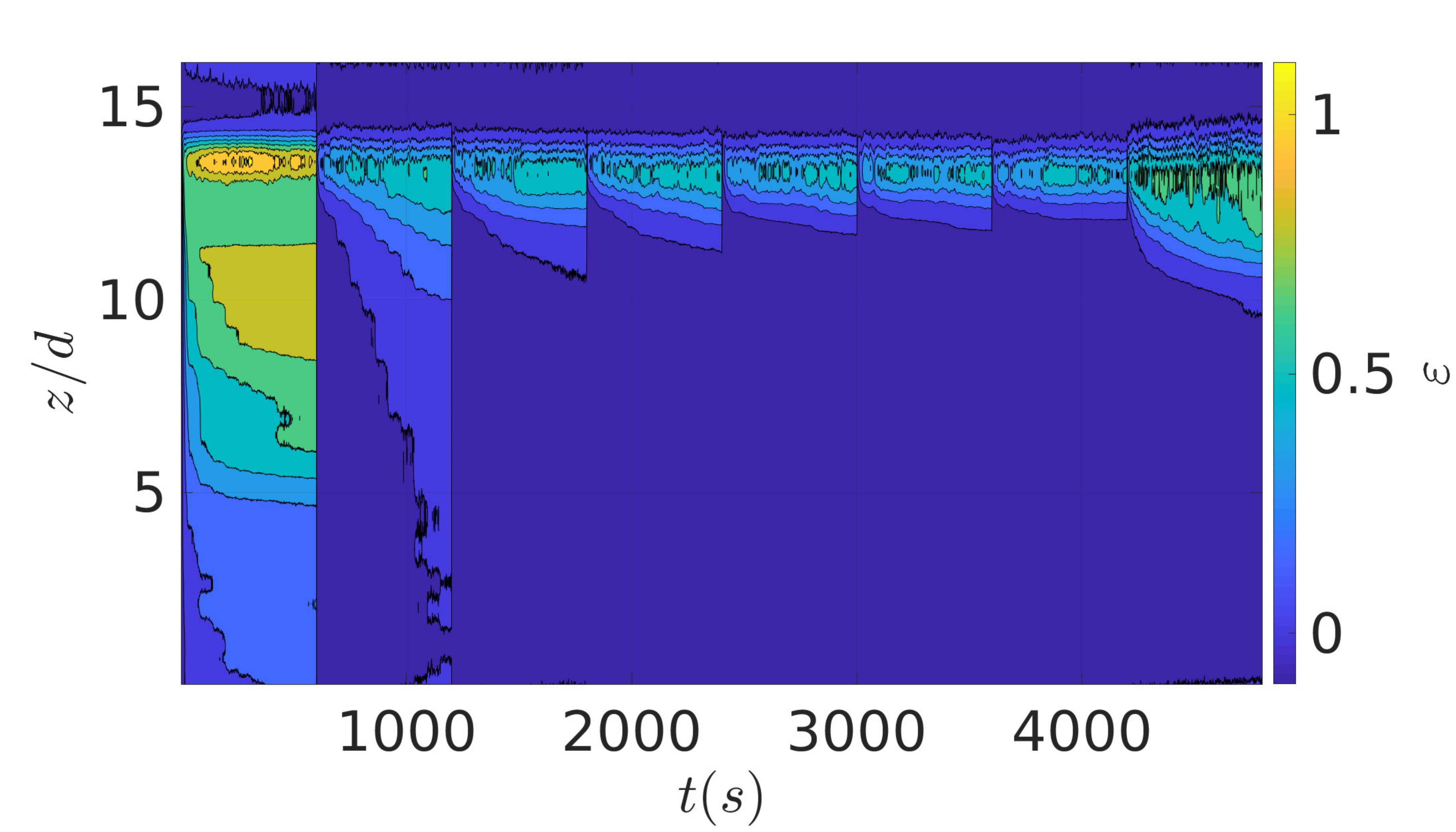}
	\caption{Same as Figure \ref{dif1}, but for test 6 (all cycles included)}
	\label{dif2_double}
\end{figure}

\begin{figure}[b]
	\includegraphics[width=0.8\linewidth]{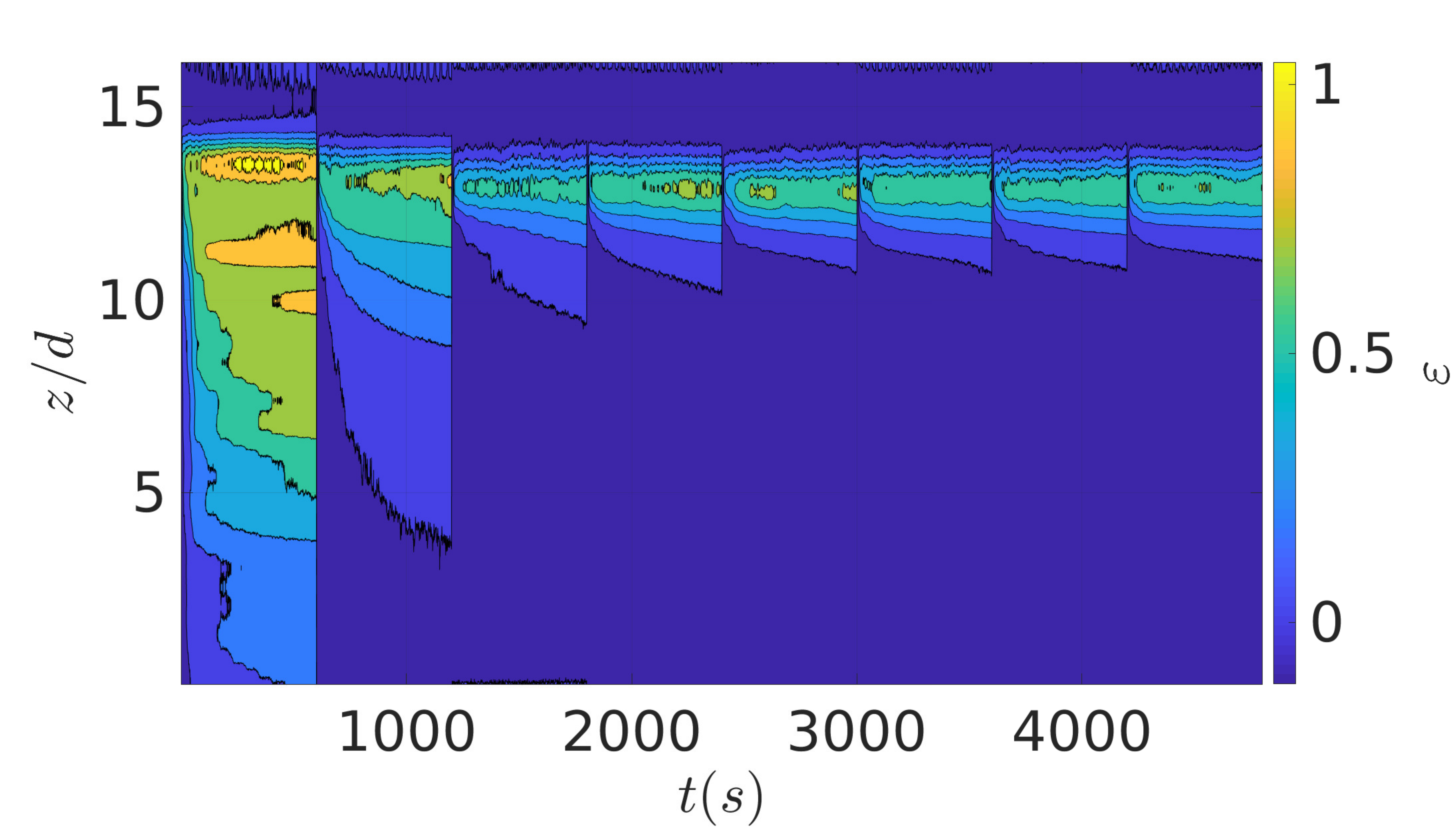}
	\caption{Same as Figure \ref{dif1}, but for test 8 (all cycles included)}
	\label{dif2_osc}
\end{figure}

\end{document}